\documentclass[amssymb,amsmath,prd,twocolumn,preprintnumbers,superscriptaddress,nofootinbib]{revtex4-1}

\usepackage{graphicx}
\usepackage{bm}
\usepackage{amssymb, amsmath, amsfonts}
\usepackage{mathrsfs}
\usepackage{latexsym}
\usepackage{color}
\usepackage[normalem]{ulem} 
\usepackage{dcolumn}
\usepackage[colorlinks=true,citecolor=blue,urlcolor=blue]{hyperref}
\usepackage[usenames,dvipsnames]{xcolor}
\usepackage{xspace}
\usepackage{graphicx}
\usepackage{booktabs}
\usepackage{siunitx}
\usepackage{nccmath}

\usepackage{multirow}
\usepackage{xfp} 
\usepackage{comment}
\usepackage{nameref}
\usepackage{orcidlink}

\newcommand{\chiEFT}{\raisebox{0.2ex}{\ensuremath{\chi}}\text{EFT}}
\newcommand{\nsat}{n_0}

\newcommand{\MTOV}{\ensuremath{M_{\rm TOV}}}

\DeclareMathOperator{\M}{M}

\usepackage[english]{babel}
\usepackage{xspace}
\usepackage{color,soul}

\newcommand{\CIT}{\affiliation{TAPIR, California Institute of Technology, Pasadena, CA 91125, USA}}
\newcommand{\CITLab}{\affiliation{LIGO Laboratory, California Institute of Technology, Pasadena, California 91125, USA}}

\begin{document}


\title{Nonparametric extensions of nuclear equations of state: probing the breakdown scale of relativistic mean-field theory}

\author{Isaac Legred\,\orcidlink{0000-0002-9523-9617}}
\email{ilegred@caltech.edu}
\CIT \CITLab

 \author{Liam Brodie\,\orcidlink{0000-0001-7708-2073
}}
 \email{b.liam@wustl.edu}
 \affiliation{Department of Physics, Washington University in St.~Louis, St.~Louis, MO 63130, USA}

 \author{Alexander Haber\,\orcidlink{0000-0002-5511-9565}}
 \email{ahaber@physics.wustl.edu}
 \affiliation{Mathematical Sciences and STAG Research Centre, University of Southampton, Southampton SO17 1BJ, United Kingdom}
 \affiliation{Department of Physics, Washington University in St.~Louis, St.~Louis, MO 63130, USA}

\author{Reed Essick\,\orcidlink{0000-0001-8196-9267}}
\email{essick@cita.utoronto.ca}
\affiliation{Canadian Institute for Theoretical Astrophysics, 60 St.~George St, Toronto, Ontario M5S 3H8}
\affiliation{Department of Physics, University of Toronto, 60 St.~George Street, Toronto, ON M5S 1A7}
\affiliation{David A.~Dunlap Department of Astronomy, University of Toronto, 50 St.~George Street, Toronto, ON M5S 3H4}

\author{Katerina Chatziioannou\,\orcidlink{0000-0002-5833-413X}}
\email{kchatziioannou@caltech.edu}
\CIT \CITLab

\begin{abstract}

Phenomenological calculations of the properties of dense matter, such as relativistic mean-field theories, represent a pathway to predicting the microscopic and macroscopic properties of neutron stars.  However, such theories do not generically have well-controlled uncertainties and may break down within neutron stars.  To faithfully represent the uncertainty in this breakdown scale, we develop a hybrid representation of the dense-matter equation of state, which assumes the form of a relativistic mean-field theory at low densities, while remaining agnostic to any nuclear theory at high densities.  To achieve this,  we use a nonparametric equation of state model to incorporate the correlations of the underlying relativistic mean-field theory equation of state at low pressures and transition to more flexible correlations above some chosen pressure scale.  We perform astrophysical inference under various choices of the transition pressure between the theory-informed and theory-agnostic models.  We further study whether the chosen relativistic mean-field theory breaks down above some particular pressure and find no such evidence.  Using simulated data for future astrophysical observations at about two-to-three times the precision of current constraints, we show that our method can identify the breakdown pressure associated with a potential strong phase transition.    
\end{abstract}

\maketitle
\section{Introduction}

Uncovering the equation of state (EoS) of dense matter is central to understanding neutron stars, bridging nuclear physics and astrophysics~\cite{Oppenheimer:1939ne, Rhoades:1974fn, Lattimer:2000nx, Lovato:2022vgq, Chatziioannou:2024tjq}.  
While the EoS is in principle determined entirely by the standard model, in practice, there is currently no perturbative or exact approach to compute the properties of matter in the densest regions of neutron star cores.
Therefore, for the purposes of interpreting macroscopic astronomical observables such as neutron star masses and radii, ``model-agnostic" representations of the EoS minimize phenomenological model-dependence~\cite{Landry:2018prl, Essick:2019ldf,Legred:2022pyp}. 
However, fully-agnostic approaches can overestimate uncertainties in regions where calculations from models of nuclear matter are reliable, for example, near nuclear saturation density ($\nsat$), where a description of matter with nucleon degrees of freedom is applicable.  
In addition, extracting detailed microphysical information from model-agnostic approaches is not straightforward, because it is not usually clear what exact microscopic interaction is responsible for determining the EoS at different densities. 
Certain quantities are not accessible at all, such as the single-particle energies required to compute transport properties \cite{Schmitt:2017efp}.  
In contrast with model-agnostic strategies, a  variety of phenomenological approaches
have been used to model dense matter up to very high densities, but these come with varying degrees of (often difficult to quantify) systematic uncertainty. These can be broken down into two categories: (1) microscopic approaches 
such as Nambu-Jona-Lasinio models~\cite{Nambu:1961tp,Buballa:2003qv} Skyrme models~\cite{Skyrme:1961vq,RevModPhys.75.121}, relativistic mean-field theories
~\cite{Walecka1974,Dutra:2014qga}, and parity-doublet models~\cite{Detar:1988kn,Jido:2001nt}, where systematic uncertainty arises from our limited knowledge of the relevant particles and their interactions, and (2) parametric approaches such as piecewise-polytrope models~\cite{Read:2008iy} and spectral models~\cite{Lindblom:2010bb} that do not explicitly model microphysics and enforce a fixed functional form which may not reflect the true EoS~\cite{Legred:2022pyp}.

Chiral effective field theory~\cite{Weinberg:1990rz} (\chiEFT{}) provides constraints on neutron-rich matter that are systematically improvable, but calculations cannot be extended to the cores of astrophysical neutron stars~\cite{Tews:2012fj, Drischler:2020hwi, Tews:2018kmu}. 
\chiEFT's nature as an effective theory allows us to systematically write down a series of interaction terms and compute an associated truncation uncertainty. 
However, its validity is limited to densities $\lesssim2\,\nsat$, which does not extend to neutron star cores. 
Relativistic mean-field theories (RMFTs), in contrast, are usable at densities and temperatures relevant in neutron stars and their mergers~\cite{Walecka1974,Boguta:1977xi,Serot:1984ey,Glendenning1996,Dutra:2014qga}.
RMFTs are phenomenological, microphysical theories of dense matter based on meson-exchange Lagrangians. We refer to a specific Lagrangian function as a single RMFT. By specifying a set of interaction constants within the RMFT, we generate a particular RMFT model that can be used to compute the corresponding EoS.

Once the functional form of the Lagrangian and the nucleon-meson and meson-meson coupling constants are determined through fits to, e.g., properties of finite nuclei, nuclear matter, and astrophysical observations, e.g.~Refs.~\cite{Fattoyev:2010mx,Steiner:2012rk,Alford:2022bpp}, no further parameters are needed to obtain finite-temperature~\cite{Alford:2023rgp} or out-of-equilibrium~\cite{Alford:2024xfb,Alford:2023gxq} properties. 
The EoS obtained from an RMFT model is also thermodynamically consistent and causal because all thermodynamic quantities are consistently derived from a partition function obtained from the relativistic RMFT Lagrangian. 
An RMFT model thus allows us to relate the microscopic properties of dense matter to astrophysical observables of neutron stars, such as their masses and radii. 

Despite these major advantages, every RMFT also comes with limitations that could lead to a breakdown of the theory, i.e., where the RMFT is no longer a plausible description of nature.
RMFTs are not obtained by a controlled expansion and thus do not have an intrinsic breakdown scale and a well-defined error. 
Furthermore, they rely on the mean-field approximation and describe nucleons as point-like interacting particles.  
Thus, while an RMFT can be used at arbitrary densities, it might not describe nature above a certain scale. 
Interpreting neutron star data assuming that the whole EoS (from the star surface to the core) is described by the RMFT model can lead to systematic biases both in macroscopic and microscopic quantities.

We employ the methodology introduced in~\citet{Essick:2020flb, Essick:2021kjb, Essick:2021ezp} as a way to avoid such biases and infer the breakdown scale with astrophysical observations. 
Although we focus on a specific RMFT in this paper, this method is generic and can be applied to any nuclear framework that is expected to become less reliable with increasing density.
The key element is the flexibility to choose how model information from the RMFT EoS is used in constructing an EoS prior with which to analyze the data.
We achieve this by incorporating information from an RMFT EoS into an otherwise model-agnostic approach for representing the EoS based on Gaussian Processes (GPs).  
The means and covariances of the GP are modified such that they closely follow RMFT EoSs generated using different choices of model parameters up to some value for the pressure.
That pressure corresponds to a choice of a scale up to which we trust the RMFT to describe the properties of dense matter.
Beyond that pressure, the GP smoothly transitions from an RMFT-informed kernel to a model-agnostic one.
We construct multiple GPs for different choices of this transition pressure.

If the true EoS can be described by an RMFT at certain densities but not up to arbitrarily high densities, then we expect this to be discernible using astrophysical observations of neutron stars. 
We consider gravitational-wave observations of masses and tidal deformabilities~\cite{Abbott:2018exr, LIGOScientific:2020aai}, X-ray observations of masses and radii~\cite{Miller:2019nzo, Riley:2019yda, Miller:2021qha, Riley:2021pdl, Choudhury:2024xbk}, and pulsar timing observations of heavy stars~\cite{Antoniadis:2013pzd, Cromartie:2019kug, Fonseca:2021wxt}.  
We use a hierarchical inference scheme to infer the EoS under a variety of hypotheses about the breakdown scale of the RMFT framework.  
We consider both simulated and real astrophysical data, and study the extent to which we can infer the breakdown scale from astrophysical data alone.
We find that with current astrophysical constraints, there is no strong evidence in favor of a breakdown of the studied RMFT.
Nonetheless, we also find that a breakdown can be approximately identified using observations at around $3\times$ the currently available measurement precision if such a breakdown is associated with, e.g., a strong, first-order phase transition. 

Using an array of different astrophysical observables targeting different mass ranges is crucial.
The primary difference between modeled and model-agnostic approaches is that the former imposes significantly stronger correlations between density scales, which might restrict its ability to fit observations~\cite{Legred:2022pyp}.    
For example, an RMFT EoS might not be capable to simultaneously explain observations of stars with a typical mass of ($\sim 1.4 \,M_{\odot}$) and the highest mass ($\gtrsim 2\, M_{\odot}$) if a strong first-order phase transition causes massive stars to have large, exotic cores.  
On the other hand, the model-agnostic GP can produce EoSs with very short correlation lengths, able to model arbitrary causal and stable EoSs, including those with phase transitions~\cite{Legred:2022pyp,Essick:2023fso}. 

The rest of the paper is organized as follows. 
In Sec.~\ref{sec:rmf-gp-description}, we discuss the RMFT EoSs we use in constructing a hybrid agnostic-informed analysis, the construction of model-agnostic and RMFT-informed GP priors, and the details of the hierarchical inference scheme. 
In Sec.~\ref{sec:simulated-analysis}, we demonstrate that the constructed EoS distributions allow us to identify a breakdown of the RMFT using sufficiently many simulated astrophysical observations. 
In Sec.~\ref{sec:astro-analysis}, we discuss the current constraints.
We do not find evidence against an RMFT description for the EoS because the uncertainties are substantial. 
We conclude in Sec.~\ref{sec:discussion}.  

\section{Methods}
\label{sec:rmf-gp-description}

We now describe how we construct our hybrid RMFT-informed, model-agnostic GP prior, and infer the EoS using astrophysical data.  In Sec.~\ref{sec:constructing RMF EoS} we describe how the RMFT EoSs used in this study are generated.  In Sec.~\ref{sec:gp-eos-prior} we review the GP EoS prior.  In Sec.~\ref{sec:RMF-GP hybrid} we describe the process of building the hybrid RMFT-informed, model-agnostic GP kernel.
Finally, in Sec.~\ref{sec:hier} we discuss intricacies of model selection relevant to the breakdown scale study.

\subsection{The RMFT EoS set}
\label{sec:constructing RMF EoS}

Like any nuclear model, a specific RMFT depends on a number of ingredients. 
Firstly, if we choose a model with only nucleons and leptons as fundamental degrees of freedom, the RMFT will not show the appearance of or the transition to new degrees of freedom unless we explicitly construct a transition to a different theory or adjust the particle composition. 
For example, we can extend an existing theory with neutrons and protons to include strange baryons, delta resonances, or N(1535) resonances~\cite{Glendenning:1982nc,Detar:1988kn,Kolomeitsev:2016ptu}. 
A transition to a completely different theory is also possible, for example, to model a phase of deconfined quark matter~\cite{Brodie:2023pjw}. 
Secondly, the choice of the Lagrangian, including the meson fields that model the strong interaction, is somewhat arbitrary. 
In principle, there are an infinite number of meson interaction terms that can be included in an RMFT \cite{Serot:1997xg}. 
To avoid overfitting, we choose a widely used Lagrangian density with the least number of coupling constants (seven) that can make predictions consistent with low-energy nuclear physics and, in principle, with astrophysical observations of neutron star structure. 
We choose a Lagrangian with a functional form identical to the IU-FSU model with seven undetermined coupling constants~\cite{Fattoyev:2010mx}.

Commonly, an RMFT is further constrained by astrophysical observables.
However, we do not want to make such an assumption \textit{a priori}, as we would potentially use observations of matter that might not be comprised of nucleons at all. 
Thus, we do not require the RMFT EoSs to support $\sim 2~M_{\odot}$ stars.
Since in our hybrid approach the RMFT description only holds at low densities, the resulting hybrid EoSs might still be able to produce high mass NSs (see e.g.~Ref.~\cite{Brodie:2023pjw}).

Even in the absence of a phase transition, at densities a few times nuclear saturation density ($\nsat \approx 0.16\ \text{fm}^{-3}$), a description of dense matter in terms of point-like nucleons is no longer plausible \cite{Weber:2004kj}. 
Since the RMFT does not include, e.g., short-range correlations between the nucleons, this effect can not be captured.  
Other potential breakdown scenarios of the RMFT include the appearance of inhomogeneous phases like chiral density waves, although this can be modeled with RMFTs~\cite{Dautry:1979bk,Papadopoulos:2024agt}, or a breakdown of the mean-field approximation where approaches such as Ref.~\cite{Friman:2019ncm} may be needed. 
In this paper, we focus on nuclear matter consisting of neutrons, protons, and electrons. 
While muons could be added to the theory, they affect the total pressure at the one percent level \cite{Alford:2022bpp}.

We produce 1109 EoSs using the method described in Ref.~\cite{Alford:2022bpp}. 
All EoSs come from the same RMFT Lagrangian density with different nucleon-meson and meson-meson coupling constants. 
The variation in these seven couplings is due to a fit to the \chiEFT\ energy per nucleon uncertainty band in neutron matter. All EoSs are causal, consistent with the saturation properties of isospin-symmetric nuclear matter, and consistent with \chiEFT\ at next-to-next-to leading order (N$^2$LO) up to $1.5\,\nsat$ \cite{Tews:2018kmu}. Of the 1109 EoSs, only 90 predict neutron stars with $M\geq2\,\M_{\odot}$.  
This set serves as a fiducial representation of RMFT EoSs.  We will occasionally refer to the RMFT EoS set as a ``prior", by which we mean taking all EoSs from the set to be equally likely.   
More details are provided in Appendix~\ref{sec:rmft-eos-distribution}.

\subsection{The GP EoS prior}
\label{sec:gp-eos-prior}

The GP EoS distribution is designed to be sufficiently flexible to incorporate information from the RMFT at low pressures while representing a fully model-agnostic EoS distribution at higher pressures.  
A GP achieves this flexibility by tuning correlations in the EoS between different scales, which are controlled directly by the covariance kernel.  
As in Refs.~\cite{Landry:2018prl, Essick:2019ldf, Essick:2020flb, Essick:2021kjb, Essick:2021ezp}, we construct a prior which is a mixture of GPs on the variable 
\begin{equation}
    \phi(\log p) = \ln \left(\frac{1}{c_s^2} - 1\right) \,,
\end{equation}
where $\log p$ is the natural logarithm of the pressure and $c_s$ is the zero-temperature, beta-equilibrated sound-speed (with $c=1$).
To generate an EoS, we sample a GP from the mixture model's mixing fractions and then draw $\phi$ from the corresponding GP (in practice, we use a finite number of pressure collocation points)
\begin{equation}
    \phi(\log p_i) \sim \mathcal N(\mu_i, C_i) \, ,
\end{equation}
where $\mu_i$ and $C_i$ are the mean and covariance associated with the $i$-th component of the mixture model.  
For different choices of mean and covariance, each pair gives a prior $\pi(\epsilon | \mathcal M)$, with $\epsilon$ the EoS and $\mathcal M$ the underlying choice of model, which corresponds to different assumptions about the EoS distribution. 
The base EoS prior is ``agnostic" in the sense that we use a mixture of many different wide covariances with short correlation lengths.
While these distributions are conditioned on nuclear models~\cite{Landry:2018prl, Essick:2019ldf}, the choice of EoSs used in the conditioning does not strongly impact the EoS prior~\cite{Legred:2022pyp}.

\subsection{Constructing an RMFT-informed GP prior and hybridization with a model-agnostic GP}
\label{sec:RMF-GP hybrid}

We construct EoS priors conditioned on the RMFT EoS distribution up to several maximum (or ``transition'') pressures as was done in Refs.~\cite{Essick:2020flb, Essick:2021kjb, Essick:2021ezp}  for \chiEFT.
However, unlike \chiEFT, the RMFT is not automatically equipped with theoretical uncertainty estimates.
That is, the distribution of RMFT EoSs introduced in Sec.~\ref{sec:constructing RMF EoS} does not include any estimate of systematic uncertainty from the fact that the RMFT Lagrangian includes a subset of all the possible terms.
In contrast, \chiEFT{} systematic uncertainties are constructed from estimates of the truncation error introduced by only retaining terms up to a certain order in the EFT expansion.  Therefore, at densities greater than saturation, the RMFT EoS distribution depends on the strategy we use to generate coupling constants from fitting inferred experimental results and \emph{ab initio} calculations; see  Sec.~\ref{sec:constructing RMF EoS}.

With these caveats in mind, we adopt the distribution of RMFT EoSs from Sec.~\ref{sec:constructing RMF EoS} and extend them with flexible, model-agnostic EoS representations above the transition pressure.
The marginal pressure distribution from the RMFT EoS is significantly skew-right and, therefore, is not well-modeled by a single GP.
Instead we emulate the RMFT EoS set with a mixture of three GPs.
Specifically, we separate the RMFT EoS set into three populations based on the marginal distribution of pressures at high densities; individual RMFT EoS tend to approach a constant $c_s$ and therefore naturally separate into different populations at high densities and pressures. In particular, we take the three sets to be those EoS that satisfy $p<1.05 \times 10^{15}\,\rm{g}/\rm{cm}^3$, $1.05\times 10^{15}<p<2\times 10^{15}\,\rm{g}/\rm{cm}^3$, and $p>2.0\times 10^{15}\,\rm{g}/\rm{cm}^3$ at a reference baryon rest-mass density of $2.8\times 10^{15}\,\rm{g}/\rm{cm}^3$.
For each of these populations, we construct a separate GP using the sample mean and covariance of the RMFT EoSs.
As a result, each GP contains strong intra-density correlations which are representative of the underlying subset of the RMFT EoS distribution it emulates; these correlations enforce smoothness in the resulting EoS realizations.
A mixture over subpopulations is constructed by weighting each subpopulation's GP based on the number of RMFT EoS that belong to the corresponding subpopulation.
Compared to Refs.~\cite{Essick:2020flb, Essick:2021kjb, Essick:2021ezp}, the distribution of RMFT EoS is more complicated and therefore requires a mixture of GPs instead of a single GP.
Readers who are interested in other approaches to emulate EoS models with GPs may also be interested in Refs.~\cite{Drischler:2020hwi, Drischler:2020yad}.

Equipped with a GP emulator for the RMFT EoS distribution, we then construct priors that closely follow the RMFT EoS distribution at low pressures but transition to more flexible model-agnostic priors above a transition pressure.
References~\cite{Essick:2020flb, Essick:2021kjb, Essick:2021ezp} achieved a similar effect by conditioning a model-agnostic prior on a \chiEFT{} emulator as if the emulator was an additional (noisy) observation.
Nevertheless, we modify the approach here.
Instead of conditioning an agnostic GP on the RMFT EoS emulator as if it was an additional observation, we instead construct a GP that exactly follows the RMFT EoS emulator at low pressures and then condition the agnostic GP at higher pressures.
The new procedure guarantees that the resulting GP will only support smooth functions at low densities if the RMFT EoS emulator only supports smooth functions at low densities.
It can be shown that the two approaches are equivalent if the (co)variance in the RMFT EoS emulator is very small compared to the (co)variance in the agnostic GP.
See Appendix~\ref{sec:fix-marginal} for details.

For emulators and agnostic priors that are mixture models of GPs (with $N$ and $M$ elements, respectively), we construct a new GP for each of the $N\times M$ pairs of GPs from the emulator and agnostic processes.
We form a mixture model over all possible combinations with weights equal to the product of the individual weights within the emulator and agnostic mixtures.
See Appendix~\ref{sec:fix-marginal} for more discussion.
An implementation of this procedure is publicly available~\cite{essick_2024_13241272}.

    \begin{table}[]
        \centering
        \caption{Transition pressures for the hybrid RMFT-agnostic GPs in two different units.
        }
        \begin{tabular}{|c||c|c|c|c|c|c|}\hline 
            $p_\mathrm{t}/c^2$ [g/cm$^3$]\rule{0pt}{3ex}   \,&$10^{11}$  &$10^{12}$  & $3\times10^{12}$ & $10^{13}$ & $3\times10^{13}$  & $10^{14}$  \\
            \hline
            $p_\mathrm{t} [\rm{MeV}/\rm{fm}^3]$ &0.056& 0.56 & 1.7  & 5.6 & 17 & 56\\
            \hline 
        \end{tabular}
        \label{tab:p_transition}
    \end{table}

Using this procedure, we construct several priors that follow the RMFT EoS emulator up to different transition pressures, $p_\mathrm{t}$.  We consider six transition pressures at values given in Table~\ref{tab:p_transition}.  
The two largest transition pressures correspond to approximately the central pressures of $0.6$ and $1.4\,M_{\odot}$ stars respectively (see, e.g., Fig.~\ref{fig:rmf-injected-posteriors}'s bottom two panels.). 
For each transition pressure, we generate 30,000 samples from the GP prior, composed of three sets of 10,000 EoS each drawn from GPs conditioned on hadronic, quarkyonic, and hyperonic EoSs~\cite{Landry:2018prl, Essick:2019ldf}.  We refer to these distributions as hybrid RMFT informed-agnostic, or for brevity, hybrid RMFT-agnostic
throughout the text.  

One advantage of using a low-density model which derives from microphysics is that it is possible to extract information about the underlying nuclear physics.
We adopt the method of extracting the symmetry parameters of isospin-symmetric nuclear matter near saturation density from Refs.~\cite{Essick:2021kjb, Essick:2021ezp}.
We consider the symmetry energy at saturation $ J =S(\nsat)$, and the so-called slope of the symmetry energy $L=3\nsat dS/dn |_{\nsat}$. 
The nuclear energy per particle is expanded locally around nuclear saturation density in terms of proton-neutron asymmetry $(n_n - n_p)/(n_n+n_p) = (1-2Y_p)$ with $Y_p$ the proton fraction\footnote{The proton fraction is also (in our case) the charge fraction and the electron fraction as we assume the only charged leptons are electrons.  We do not expect the addition of muons to substantially alter the recovered parameters~\cite{Essick:2021kjb, Essick:2021ezp}. }  and the baryon number density $n$:
\begin{equation}
    \label{eq:symmetry-expansion}
    \frac{E_{\rm nuc}}{A}(n, Y_p) = \frac{E_{\rm SNM}(n)}{A} + S(n)(1-2Y_p)^2\,,
\end{equation}
where $E_{\rm SNM}$ is the energy of symmetric nuclear matter, which we additionally expand as 
\begin{equation}
    E_{\rm SNM}(n) = E_{\rm bind} + \frac{1}{2}\kappa\left(\frac{n-\nsat}{3\nsat}\right)^2\,.
\end{equation}
We sample $E_{\rm bind}$, $\kappa$, and $\nsat$ from Gaussian distributions with mean and variance fit from the RMFT EoS set and given in Table~\ref{tab:sym-matter-parameters}.  
The nuclear energy per particle is equal to the total energy minus the contribution from electrons, which is known given the baryon number density and proton fraction
\begin{equation}
    \frac{E_{\rm nuc}}{A}(n, Y_p) = \frac{\varepsilon(n, Y_p)-\varepsilon_{e}(n, Y_p)}{n}\,,
\end{equation}
where $\varepsilon$ and $\varepsilon_{e}$ are the total and electron energy densities respectively.  
In particular, if $Y_p$ is taken to be the beta-equilibrium charge fraction ($Y^{\beta}_p$), then $\varepsilon(n, Y^{\beta}_p)$ is precisely the beta-equilibrium energy density, which is computed for each GP draw.  
Therefore, if the beta-equilibrium charge fraction were known, then $E_{\rm nuc}(n,Y^{\beta}_p)$ would also be known, as would (rearranging Eq.~\eqref{eq:symmetry-expansion})
\begin{equation}
\label{eq:symmetry-energy-extracted}
   S(n) =  \left(\frac{E_{\rm nuc}}{A}(n,Y^{\beta}_p)- \frac{E_{\rm SNM}}{A}(n)\right)/(1-2Y_p^{\beta})^2\,.
\end{equation}
The beta-equilibrium charge fraction can be found from setting $\mu_n = \mu_p + \mu_e$ (assuming a zero neutrino chemical potential); see Refs.~\cite{Hebeler:2013nza, Essick:2021kjb, Essick:2021ezp} for details. 
From this we compute 
\begin{align}
    J &\equiv S(\nsat)\,,\\
    L &\equiv 3\nsat \frac{dS(n)}{dn}\Big|_{n=\nsat}\,,\\
    K_{\rm sym} &\equiv 9\nsat^2  \frac{d^2S(n)}{dn^2}\Big|_{n=\nsat}\,,
\end{align}
for each GP draw.  
We additionally verify that for our set of RMFT EoS draws, this procedure produces reasonable estimates for the (already known) symmetry parameters.

\begin{table}[]
    \centering
        \caption{Mean and standard deviation of the parameters of symmetric matter used in computing the symmetry energy and its derivatives. These represent the mean and standard deviation of the corresponding parameters from our RMFT EoS distribution. 
    }
    \begin{tabular}{|c||c|c|}
    \hline
        Parameter & $\mu$  &$\sigma$ \\
        \hline
        $\nsat\ [\rm{fm}^{-3}]$ & 0.1568 & 0.0017  \\
        \hline
         $E_{\rm bind}\ [\rm{MeV}]$ & -15.983 & 0.046 \\ \hline
         $\kappa\ [\rm{MeV]}$ & 228 & 25 \\\hline
    \end{tabular}

    \label{tab:sym-matter-parameters}
\end{table}

\subsection{The Hierarchical Inference Scheme}
\label{sec:hier}

Given an EoS prior and astrophysical data, we infer the EoS following Refs.~\cite{Essick:2019ldf, Landry:2020vaw, Chatziioannou:2020pqz}.
In this subsection, we provide a brief overview and discuss considerations in interpreting Bayes factors.
The key ingredients are 
\begin{itemize}
    \item $\pi(\epsilon | \mathcal M)$: The prior on EoS $\epsilon$, which depends upon the model $\mathcal M$.  
    \item $\mathcal L(d_i |\epsilon)$: The likelihood, which is the probability of receiving astrophysical data $d_i$ given an EoS $\epsilon$.
\end{itemize}
The posterior probability of a given EoS is
\begin{equation}
    P(\epsilon | \{d_i\}, \mathcal M) = \frac{\prod_i \mathcal L(d_i | \epsilon)\pi(\epsilon | \mathcal M)}{\mathcal Z (\{d_i\} | \mathcal M)}\,,
\end{equation}
where ${d_i}$ represents the set of all observed data, and
 \begin{equation}
        Z(\{d_i\}_i|\mathcal M) =  \int \pi(\epsilon | \mathcal M) \prod_i \mathcal L(d_i | \epsilon) d\epsilon\,, \
\end{equation} 
is the evidence for model $\mathcal M$.
In this work $d_i$ could stand for data resulting from X-ray pulse profile measurements of neutron star mass-radii, pulsar timing observations of heavy neutron star masses, and gravitational-wave observations of neutron star masses and tidal deformabilities.  
Given two models $\mathcal M_1$ and $\mathcal M_2$, the Bayes factor compares their relative fit to the data 
\begin{equation}
    \mathcal B^{\mathcal M_1}_{\mathcal M_2} = \frac{\mathcal Z(\{d_i\} | \mathcal M_1)}{\mathcal Z(\{d_i\} | \mathcal M_2)} \, .
\end{equation}
The Bayes factor is the average likelihood of one model relative to another, where the average is taken over the entire prior volume of each model.  
A model whose prior includes regions with a low likelihood will have a low Bayes factor relative to a model that excludes such regions.
Nonparametric, model-agnostic models are explicitly designed to explore such regions in the name of flexibility.
Indeed, most samples from a model-agnostic GP prior will have very low likelihood.  
To minimize this effect, we compute Bayes factors after conditioning the EoS prior distributions on heavy pulsar observations.
This effectively removes EoSs that have too low a maximum mass, and provides a proxy for only including ``astrophysically plausible" EoSs within the prior.  

Excluding these EoSs prevents conclusions from being driven mainly by EoSs with masses much less than $2\,M_{\odot}$, instead targeting the likelihood of astrophysically plausible EoSs sampled from each model~\cite{Essick:2020flb, Legred:2021, Essick:2023fso, Mroczek:2023zxo}.

\section{Verification with simulated astrophysical observations}
\label{sec:simulated-analysis}

We demonstrate our methodology by using simulated astrophysical observations to infer the EoS, and in tandem, infer the breakdown scale of the underlying RMFT.

\subsection{Simulated Data}
\label{sec:description-of-inference}
    
We verify the hybrid RMFT-agnostic GPs with simulated data.  
To do this, we generate simulated observations using two EoSs. 
The first EoS is consistent with the distribution of RMFT EoSs at all density scales~\cite{Alford:2022bpp}; its saturation parameters are listed in Table~\ref{tab:saturation_properties}. 
The second EoS is constructed to be inconsistent with the RMFT EoSs.
At an energy density of $\varepsilon/c^2 = 1.7 m_{N} \nsat \approx 260\,\mathrm{MeV}/\rm{fm}^3 $, a strong phase transition with a latent heat of $\Delta \varepsilon/ \varepsilon = 0.4$ is inserted, after which there is a constant speed of sound $c_s^2= 0.8$.  This transition happens at a baryon density of about $1.5\,\nsat$, and a pressure of $\sim 12 \,\rm{MeV}/\rm{fm}^3$.

\begin{table}[]
  \centering
    \caption{The saturation and symmetry parameters of nuclear matter for the RMFT EoS we use to generate simulated data.  $E_B$ is the binding energy per nucleon in isospin-symmetric nuclear matter at saturation density, $\nsat$; $\kappa$ is the incompressibility of nuclear matter at $\nsat$; $J$ is the symmetry energy, the difference between the binding energy per nucleon of neutron matter and isospin-symmetric nuclear matter, evaluated at $\nsat$; $L$ characterizes how the symmetry energy varies with density; $m^*$ is the Dirac in-medium nucleon mass at $\nsat$.}
  \begin{tabular}{|c|c|c|c|c|c|}
    \hline
    $\nsat$ [fm$^{-3}$] & $E_B$ [MeV] & $\kappa$ [MeV] & $J$ [MeV] & $L$ [MeV] & $m^*$ [MeV] \\ \hline
    0.158 & \multicolumn{1}{c|}{-16} & \multicolumn{1}{c|}{244} & \multicolumn{1}{c|}{34} & \multicolumn{1}{c|}{52.9} & \multicolumn{1}{c|}{685} \\ \hline
  \end{tabular}

  \label{tab:saturation_properties}
\end{table}

We generate simulated data that represent potential observations via radio pulsar timing, x-ray pulse-profile modeling, and gravitational-wave observations.  
Starting with radio timing, we assume that pulsars are formed with masses up to the Tolman-Oppenheimer-Volkoff (TOV) maximum mass~\cite{Oppenheimer:1939ne, Tolman:1939jz}, and the mass distribution is uniform. 
We simulate radio timing observations~\cite{Shapiro:1964uw, Blanford:1976}, assuming a factor of $\sim 2$ reduction in uncertainty compared to current constraints~\cite{Demorest:2010bx, Antoniadis:2013pzd, Cromartie:2019kug}.
See Table~\ref{tab:ns_observations_rmft} for the masses and uncertainties for the RMFT EoS and Table~\ref{tab:ns_observations_pt} for the phase-transition case.
In particular, we assume two heavy pulsars with well-measured masses.

For X-ray data, we simulate 4 sources.  
We assume that uncertainties are uncorrelated and Gaussian on the mass and radius, which is reasonable for sources with independently measured masses from radio timing, such as J0740+6620~\cite{Miller:2021qha, Riley:2021pdl} and J0437-4715~\cite{Choudhury:2024xbk}.
This assumption further improves as more photon counts and better background estimates are included, ~\cite{Dittmann:2024mbo, Salmi:2024aum}, indicating that for future observations nearly-Gaussian uncertainties are plausible.  
We sample the uncertainties on mass and radius from  uniform distributions of width $.3\,M_{\odot}$ and $1.0\,\rm{km}$ respectively. 
The uncertainties in the most optimistic cases are $\sim 2$-$3\times$ smaller than current measurements.  
See Table~\ref{tab:ns_observations_rmft} for the the simulated X-ray sources for the EoS consistent with the RMFT and Table~\ref{tab:ns_observations_pt} for the EoS with a phase transition.

\begin{table}[h!]
    \centering
    \caption{Simulated neutron star observations for the RMFT EoS.  Values given represent mean and 90\% credible intervals.}
    \label{tab:ns_observations_rmft}
    \begin{tabular}{|c||c|c|}
        \toprule
        \hline
        Neutron Star & {Mass ($M_\odot$)} & {Radius (km)} \\
        \midrule
        \hline
        Radio 0  & $2.10 \pm 0.03$ & NA \\
        \hline
        Radio 1  & $2.08 \pm 0.13$  & NA \\
        \hline
        \hline

        \midrule
        X-ray 0 & $1.20 \pm 0.04$ & $12.29 \pm 1.25$ \\
        \hline
        X-ray 1 & $1.37 \pm 0.11$ & $12.26 \pm 0.25$ \\
        \hline
        X-ray 2 & $1.45 \pm 0.45$ & $12.24 \pm 0.69$ \\
        \hline
        X-ray 3 & $1.96 \pm 0.43$ & $11.77 \pm 1.10$ \\
        \hline
        \bottomrule
    \end{tabular}
\end{table}

\begin{table}[h!]
    \centering
    \caption{Simulated neutron star observations for the phase-transition EoS, values given represent mean and 90\% credible intervals. }
    \label{tab:ns_observations_pt}
    \begin{tabular}{|c||c|c|}
        \toprule
        \hline
        Neutron Star & {Mass ($M_\odot$)} & {Radius (km)} \\
        \hline
        \midrule
        Radio 0 & $2.42 \pm 0.06$ & NA \\
        \hline
        Radio 1  & $2.44 \pm 0.07$ & NA \\
        \hline 
        \hline
        \midrule
        X-ray 0  & $1.20 \pm 0.35$ & $11.63 \pm 1.08$ \\
        \hline
        X-ray 1  & $1.37 \pm 0.02$ & $11.70 \pm 0.15$ \\
        \hline
        X-ray 2  & $1.45 \pm 0.43$ & $11.73 \pm 0.94$ \\
        \hline
        X-ray 3  & $1.96 \pm 0.00$ & $11.93 \pm 1.31$ \\
        \bottomrule
        \hline
    \end{tabular}
\end{table}

For gravitational waves, we perform full parameter estimation, because even at moderate  signal-to-noise ratio  (SNR) it is empirically not straightforward to construct an analytic expression for the posterior distribution on the relevant parameters (\emph{e.g.} the distributions on the effective tidal deformability and mass ratio, $\tilde \Lambda$ and $q$ are not approximately Gaussian~\cite{gw170817, gw190425}).
We recover posterior distributions on the quantities $m_1, m_2, \Lambda_1, \Lambda_2$, the primary and secondary masses, and primary and secondary tidal deformabilities, respectively.  
We then choose 2 sources sampled randomly from the set of recovered sources.  
We display these parameters for the RMFT EoS injection in Table~\ref{tab:rmft-gw-simulations}.  
The second source is very similar to GW170817, both in terms of parameters and in terms of measurement precision.  
We display the corresponding parameters for the RMFT-PT EoS injections in Table~\ref{tab:rmft-pt-gw-simulations}.
Further gravitational-wave injections are used in Appendix~\ref{sec:simulated-inference-gws}, which also includes a detailed description of simulation study.

\begin{table}[h]
    \centering
        \caption{Simulated gravitational wave parameters including mass ($m_1$ and $m_2$), tidal deformability ($\Lambda_1$ and $\Lambda_2$), and SNR, for the RMFT EoS.}

    \begin{tabular}{|c||c|c|c|c|c|}
        \toprule
        \hline
        Binary & $m_1 \, [M_\odot]$ & $m_2\, [M_\odot]$ & $\Lambda_1$ &  $\Lambda_2$ & SNR \\
        \midrule
          \hline
       GW 0 & 1.5 & 1.18 & 266.6 & 1139.45 & 15.0\\
         \hline
       GW 1 & 1.49 & 1.17 & 273.1 & 1189.81 & 32.5\\
         \hline
        \bottomrule
    \end{tabular}
    \label{tab:rmft-gw-simulations}
\end{table}

\begin{table}[h]
    \centering
    \caption{Simulated gravitational wave parameters from the phase-transition EoS.  Same as Table~\ref{tab:rmft-gw-simulations}.}
    \begin{tabular}{|c||c|c|c|c|c|}
        \toprule
        \hline
        Binary & $m_1 \, [M_\odot]$ & $m_2\, [M_\odot]$ &  $\Lambda_1$ &  $\Lambda_2$ & SNR \\
        \midrule
        \hline
       GW 0 & 1.26 & 1.09 & 473.89 & 1011.61 & 10.7\\
         \hline
      GW 1 & 1.23 & 1.2 & 542.88 & 613.32 & 26.5\\
         \hline
    \end{tabular}

    \label{tab:rmft-pt-gw-simulations}
\end{table}

\subsection{Simulated inference results}
\subsubsection{Recovering an RMFT EoS}
\label{sec:rmf-recovery}
    
\begin{figure*}
        \centering
        
       \includegraphics[width=0.49\linewidth]{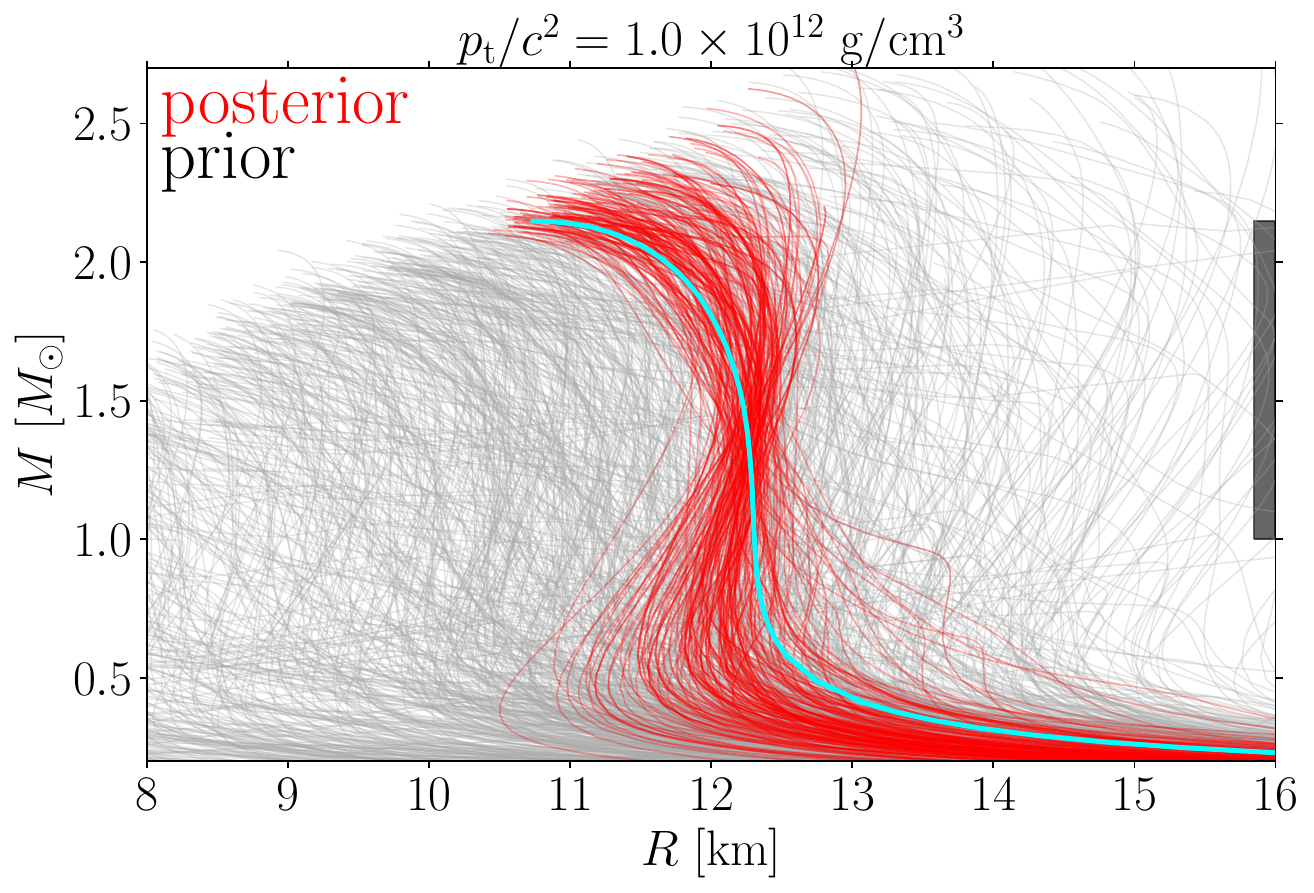}
       \includegraphics[width=0.49\linewidth]{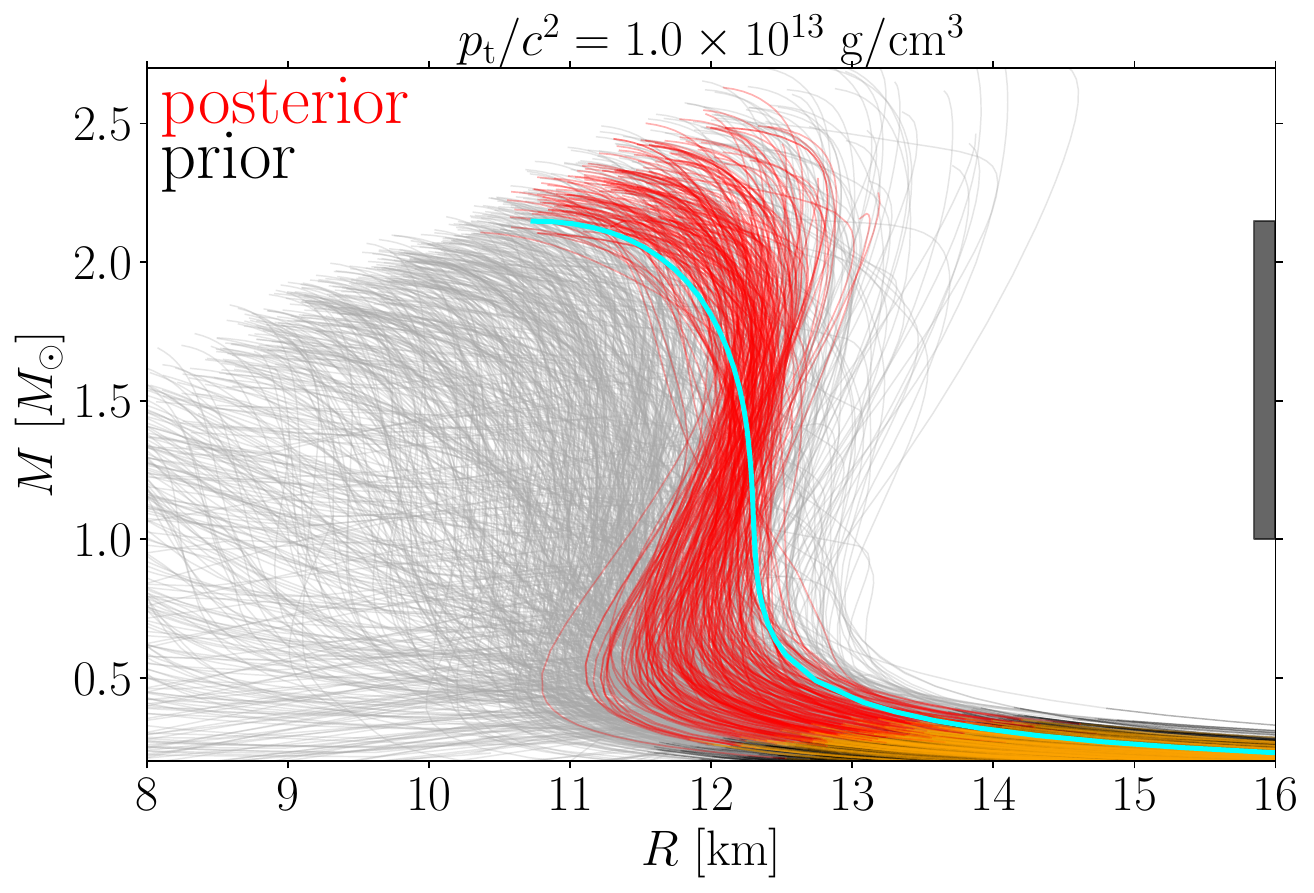}
        \includegraphics[width=0.49\linewidth]{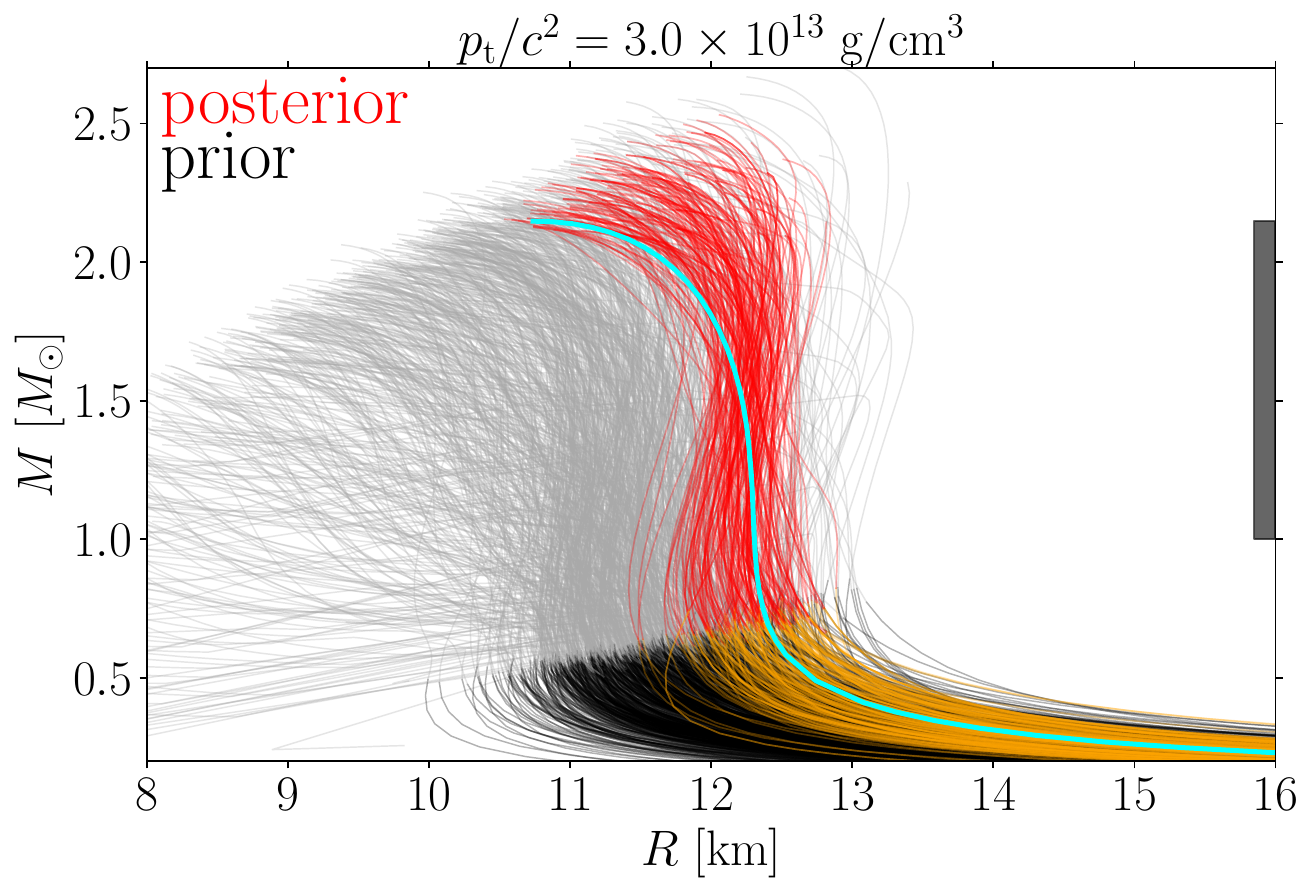}
        \includegraphics[width=0.49\linewidth]{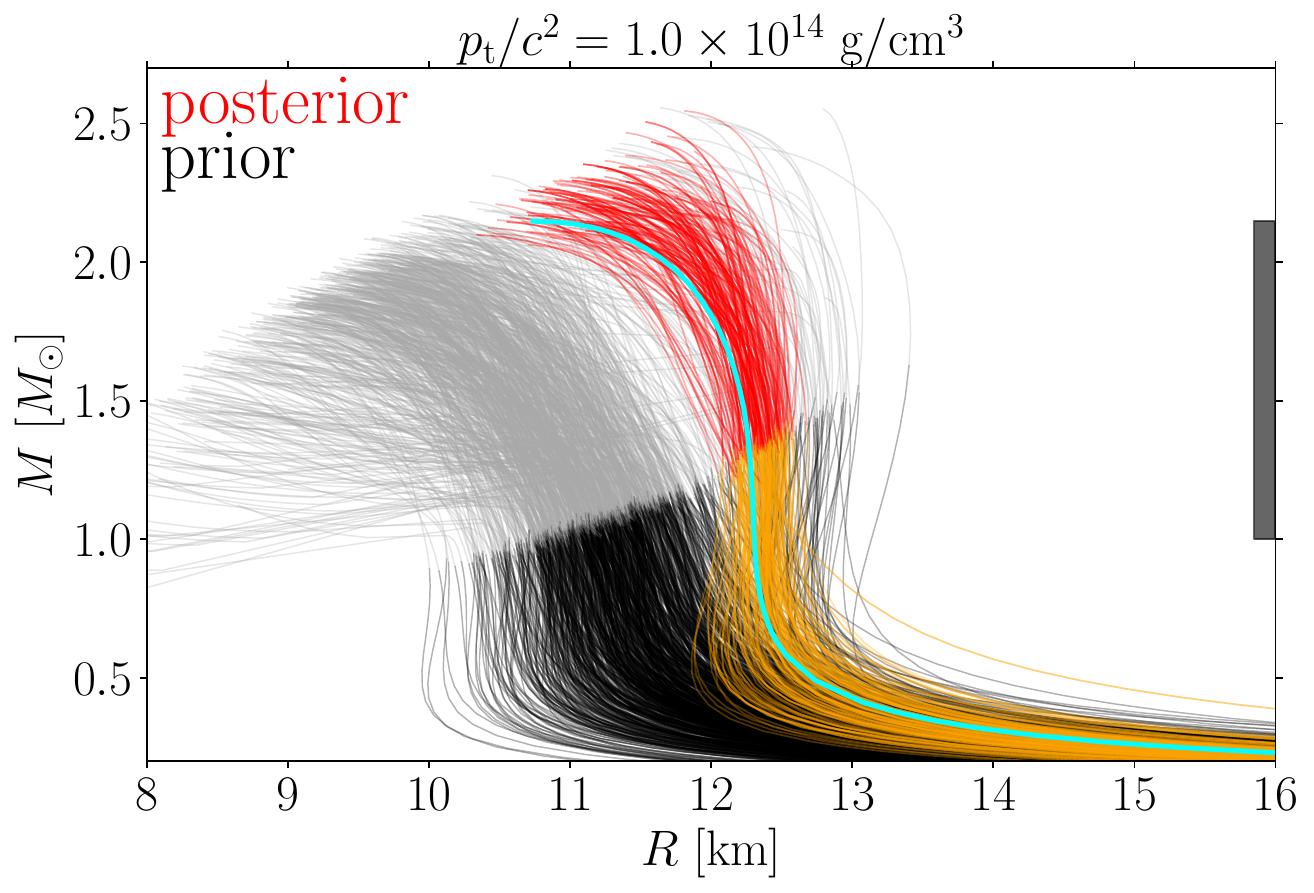}
 
        \caption{In gray/black, we show $M$-$R$ curves of fair draw EoS from RMFT-agnostic hybrid prior distributions which transition at various densities, shown by subplot titles. Stars fully informed by the RMFT are shown in black, while stars with cores that have transitioned to the agnostic model are shown in gray.  In cyan, we display the EoS used to generate simulated observations. In red/orange, we show samples from the posterior.
        In this case, points in orange are stars fully described by the RMFT, while points in red represent stars with cores that have transitioned to the agnostic model.  On the right y-axis of each panel, we display a gray bar from $M\in(M_{\odot}, M_{\rm{TOV}}) $ where $M_{\rm{TOV}}$ is the TOV maximum mass of the simulation EoS. This represents the approximate range of NS masses used in astrophysical inference.
        }
        \label{fig:rmf-injected-posteriors}
        \end{figure*}

We begin with the case of an EoS that is consistent with the RMFT EoSs.
We display the RMFT EoS used for the simulations and 200 fair draws from the posterior for prior distributions at four values of the transition pressures in Fig.~\ref{fig:rmf-injected-posteriors}.   
We additionally plot 1000 fair draws from each prior distribution. 
First, looking at the prior distributions, as more information from the underlying RMFT model is included, moving from lower to higher transition pressure (left to right and top to bottom panels), the EoS distribution is much smoother and narrower below the transition pressure.  
Conversely, above the transition, the prior is wider and less regular, with rapid changes to the radius with mass being much more common.

Moving on to the posteriors,  for all transition pressures, the simulated EoS (light blue) is recovered correctly, with an uncertainty that increases for lower transition pressures.  
This is expected because priors with lower transition pressure incorporate less information from the underlying RMFT and its strong model constraints.  
This effect is most severe where the simulated astrophysical observations are least informative, for example, at very high and low neutron star masses.  
The uncertainty in radius of a $2\,M_{\odot}$, for instance, is $\sim 1.25~\rm{km}$ with the model agnostic prior at $90 \%$ credibility.  

\begin{figure}
        \centering
               \includegraphics[width=0.49\textwidth]{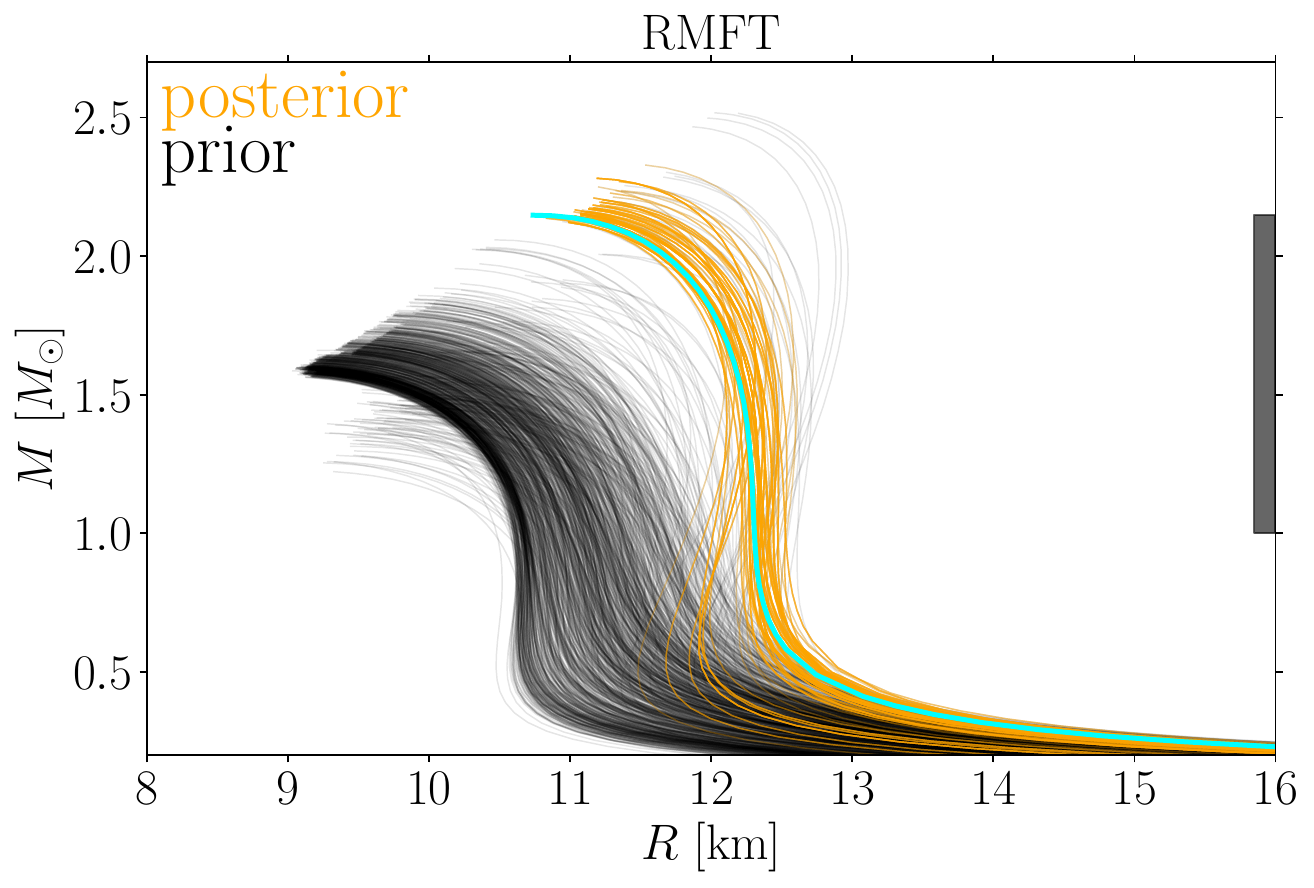}
        \caption{Inference with the RMFT prior itself.  Same as Fig.~\ref{fig:rmf-injected-posteriors}, but with the set of RMFT EoS samples with uniform probability used as prior distribution.  Since the RMFT EoSs are informed by the RMFT at all densities, we color the posterior orange and the prior black for all neutron stars, in analogy with Fig.~\ref{fig:rmf-injected-posteriors}.}
        \label{fig:rmf-inj-rmf-posterior}
    \end{figure}

For comparison, we analyze the same data with the RMFT EoS prior set itself and display the results in Fig.~\ref{fig:rmf-inj-rmf-posterior}. 
Comparing to Fig.~\ref{fig:rmf-injected-posteriors}, and in particular the bottom right panel which shows the distribution that carries the most RMFT information, we see that the overall structure of the RMFT EoS distribution is well captured by the GP emulator.  
This includes a ``bimodality" in the mass-radius relation distribution for the RMFT EoS set. 
This bimodality arises because of the choice of initial fitting parameters, as multiple combinations of the RMFT parameters are able to effectively reproduce the properties of symmetric and pure neutron matter, which have very different behavior at high density; see Appendix~\ref{sec:rmft-eos-distribution}.

\begin{figure}
        \centering
        \includegraphics[width=\linewidth]{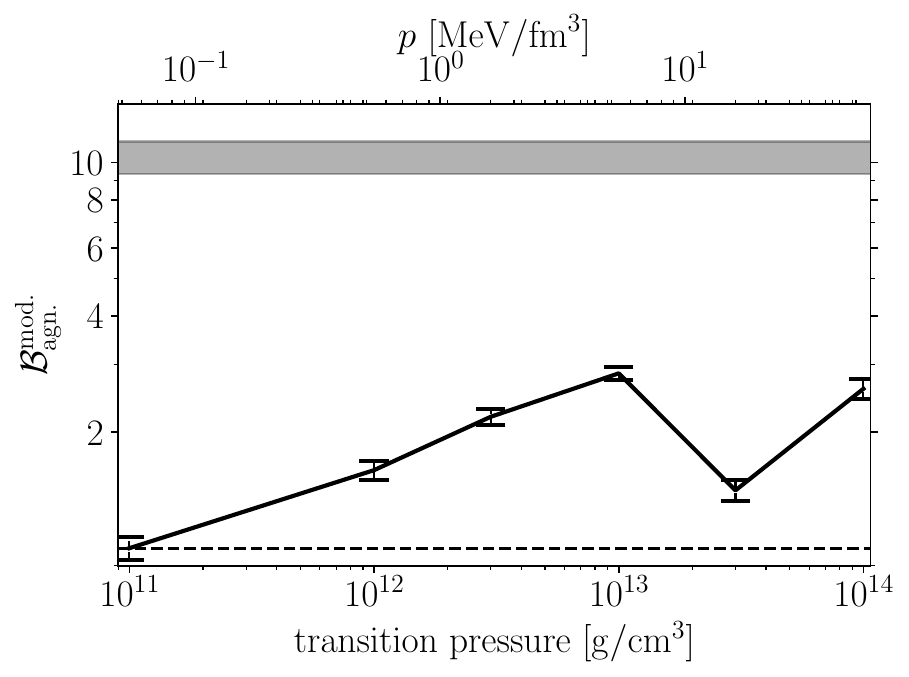}
        \caption{The Bayes factor of hybrid RMFT-agnostic GPs with various transition pressures compared to the ``agnostic model'' with a transition pressure of $10^{11} \rm{g}/\rm {cm}^3$. Points show Monte-Carlo estimates, with error bars showing $\pm 1$-$\sigma$ error in the estimate from the limited sample size of the EoSs. The 1-$\sigma$ region for the Bayes factor of the RMFT prior itself relative to the agnostic model is shown as a gray bar. There is an overall trend toward higher transition densities, but there is no conclusive preference for any transition pressure.  The simulated data are consistent with all hybrid priors.
        }
        \label{fig:rmf-bayes-factors}
    \end{figure}

For each transition pressure, $p_{\rm t}$, we compute the evidence $\mathcal Z(\{d_i\} |  \mathcal M=p_{\rm t}  )$.
This is to say, each GP conditioned at a different transition pressure forms a model $\mathcal M$, and we take a transition pressure of $p/c^2 = 10^{11} \rm{g}/\rm{cm}^3$ as a fiducial ``agnostic value" since this incorporates the least information from the underlying RMFT. 
We then compute each model's Bayes factor relative to the agnostic model and plot them in Fig.~\ref{fig:rmf-bayes-factors}.
There is a general preference for transitioning from RMFT to the more flexible GP at higher pressure, but it is weak with Bayes factors of $2$-$4$. 
Under the RMFT EoS prior, the Bayes factor relative to the agnostic model is higher, $\sim 11$.  
This is a consequence of our choice to condition the Bayes factors on the existence of heavy pulsars.  
EoSs with $M_{\rm TOV}\gtrsim 2.0$ effectively all appear in the posterior.  
Therefore, conditioned on the existence of heavy pulsars, the RMFT is more highly preferred than any agnostic model.  
In contrast, at $p_{\rm t}/c^2 = 3\times 10^{13} \, \rm{g}/\rm{cm}^3$  there is a dip in the evidence.  
This is because at this transition pressure, EoSs with low values of $R_{1.4}\sim 11.5\,\rm{km}$, which are inconsistent with the radius of the injected EoS ($R_{1.4} \sim 12.5\,\rm{km}$) are able to stiffen to reach $M_{\rm TOV} \gtrsim 2.1\,M_{\odot}$.
Therefore low-radius EoSs are common even after conditioning on the existence of massive pulsars in this case, and their marginal likelihood is lower.
If we remove this conditioning, then the RMFT is no longer substantially preferred relative to the agnostic model, and conditioning at $p_t/c^2 = 10^{14}\, \rm{g}/\rm{cm}^3$ is actually disfavored relative to the agnostic model despite the fact that the simulation EoS is itself an RMFT EoS.

The above discussion highlights the large dependence of Bayes Factors  on relatively minor analysis choices, which is why we deem Bayes Factors of ${\mathcal{O}}(10)$ as inconclusive.
Nonetheless, the features of Fig.~\ref{fig:rmf-bayes-factors} are still meaningful; for example, the trend of increasing Bayes factor with transition pressure can be attributed to removing prior volume that is inconsistent with the RMFT, and will be assigned low-likelihood.
If two models can describe the data equally well but one has less prior volume than the other (e.g., a model with a low transition pressure vs.~a model whose prior is more tightly concentrated around the RMFT), the former is preferred.  
The Bayes factor for such a preference, though, is a fixed number, even in the limit of infinite measurement precision, since it is determined by the ratio of the probability density functions of each of the EoS priors at the simulation EoS.  
Therefore, depending on what this value is, we may or may not ever have decisive evidence in favor of the RMFT over the agnostic model.

\begin{figure}
        \centering
        \includegraphics[width=0.49\textwidth]{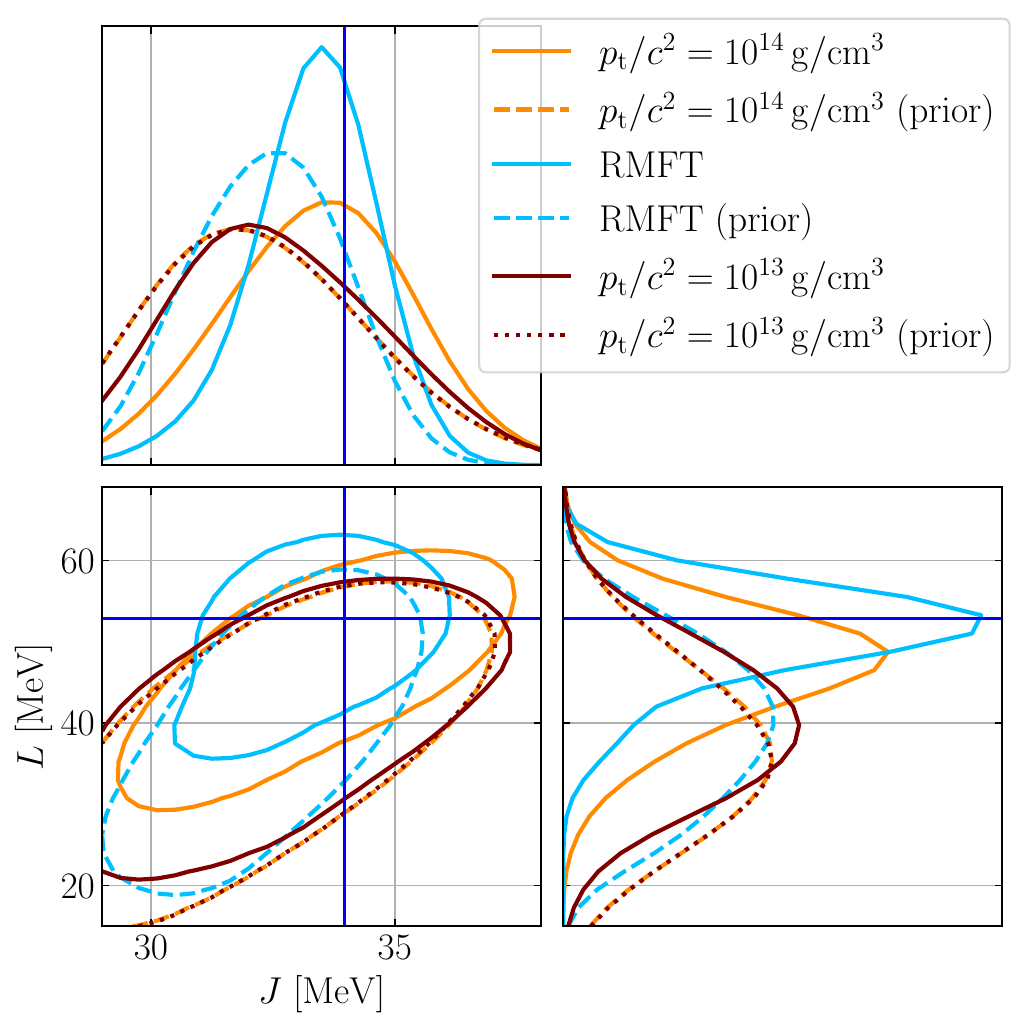}
        \caption{Prior (dashed) and posterior (solid) for the symmetry energy $L$ and the slope of the symmetry energy $J$ at saturation for various EoS priors. 
        In light blue is the result from using the set of RMFT EoSs directly as a prior. 
        In maroon and orange are results on the symmetry parameters (as estimated using  Eq.~\ref{eq:symmetry-energy-extracted}) inferred using GP-priors which transitioned from RMFT-informed to model agnostic kernels at $10^{13}$ and $10^{14}\,\rm{g}/\rm{cm}^3$ respectively.  For the GP EoS distributions ($p_{\rm t}/c^2
        =10^{13}$ and $p_{\rm t}/c^2=10^{14}\,\rm{g}/\rm{cm}^3$), the prior distributions are effectively identical, since both follow the same RMFT-informed GP at saturation density.  Therefore, we mark the prior for the $p_{\rm t}/c^2 = 10^{13}\,\rm{g}/\rm{cm}^3$  with a dotted line to increase visibility.
        }
        \label{fig:rmf-symmetry-params}
    \end{figure}

    Finally, we consider the symmetry energy parameters.  
    We compute the posterior for $J$ and $L$, the symmetry energy at saturation and the slope of the symmetry energy at saturation, respectively, and display the results in Fig.~\ref{fig:rmf-symmetry-params}.
    The true symmetry parameters are comfortably recovered at 90\% credibility regardless of the choice of  transition pressure. As we increase the transition pressure, the prior and posterior become narrower and approach the result of using the RMFT EoS set itself.  
    This is consistent with strengthening correlations between symmetry parameters and astrophysical observables as the RMFT is trusted to higher and higher pressures. 
    These posteriors all incorporate the same data and have very similar marginal priors on the symmetry parameters.
    Therefore differences between them are driven solely by differences in the higher-dimensional EoS prior distributions, as also observed in Refs.~\cite{Essick:2021kjb,Essick:2021ezp,Legred:2022pyp}.

\subsubsection{Inferring a phase-transition EoS}
\label{sec:pt-recovery}

\begin{figure*}
     \centering
     \includegraphics[width=0.49\linewidth]{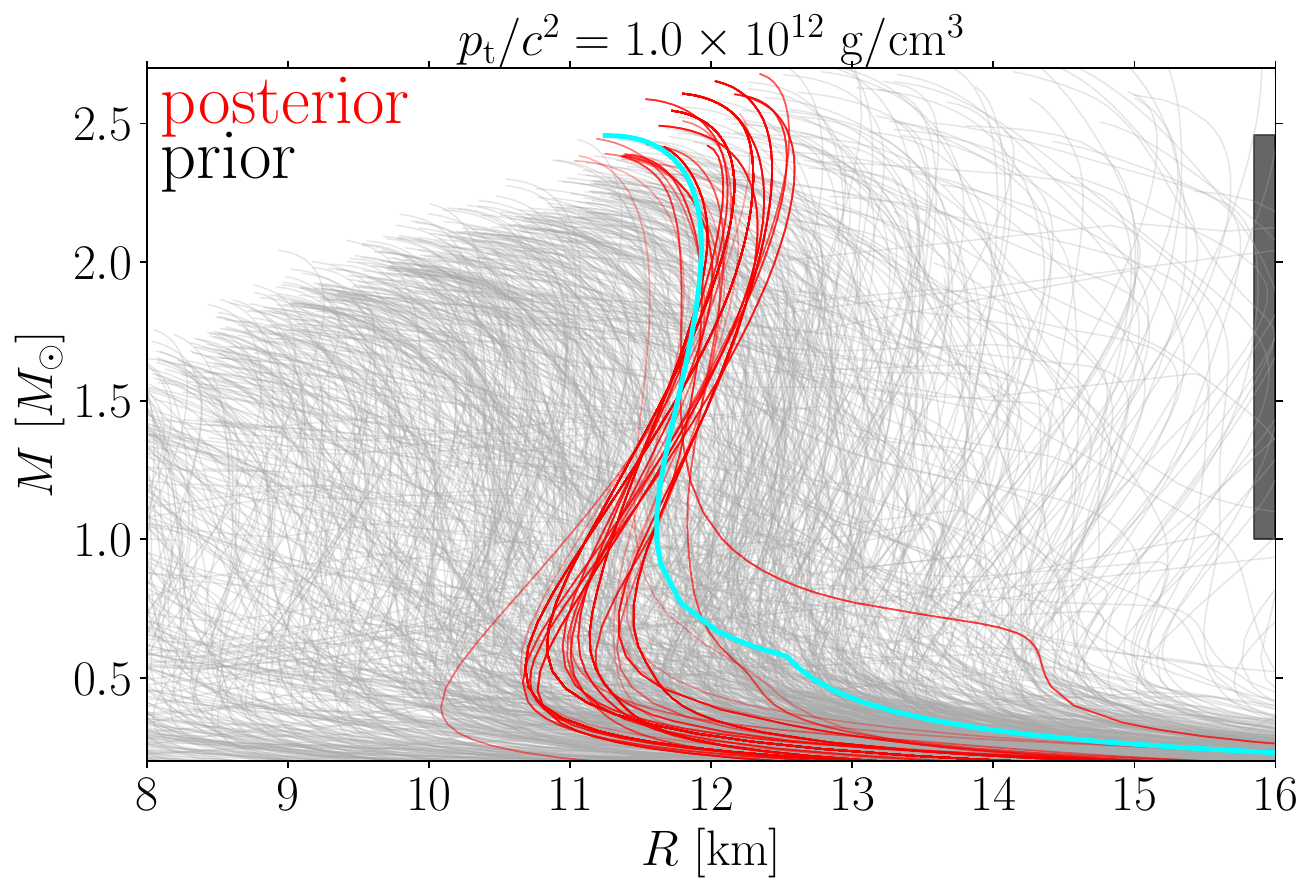}
       \includegraphics[width=0.49\linewidth]{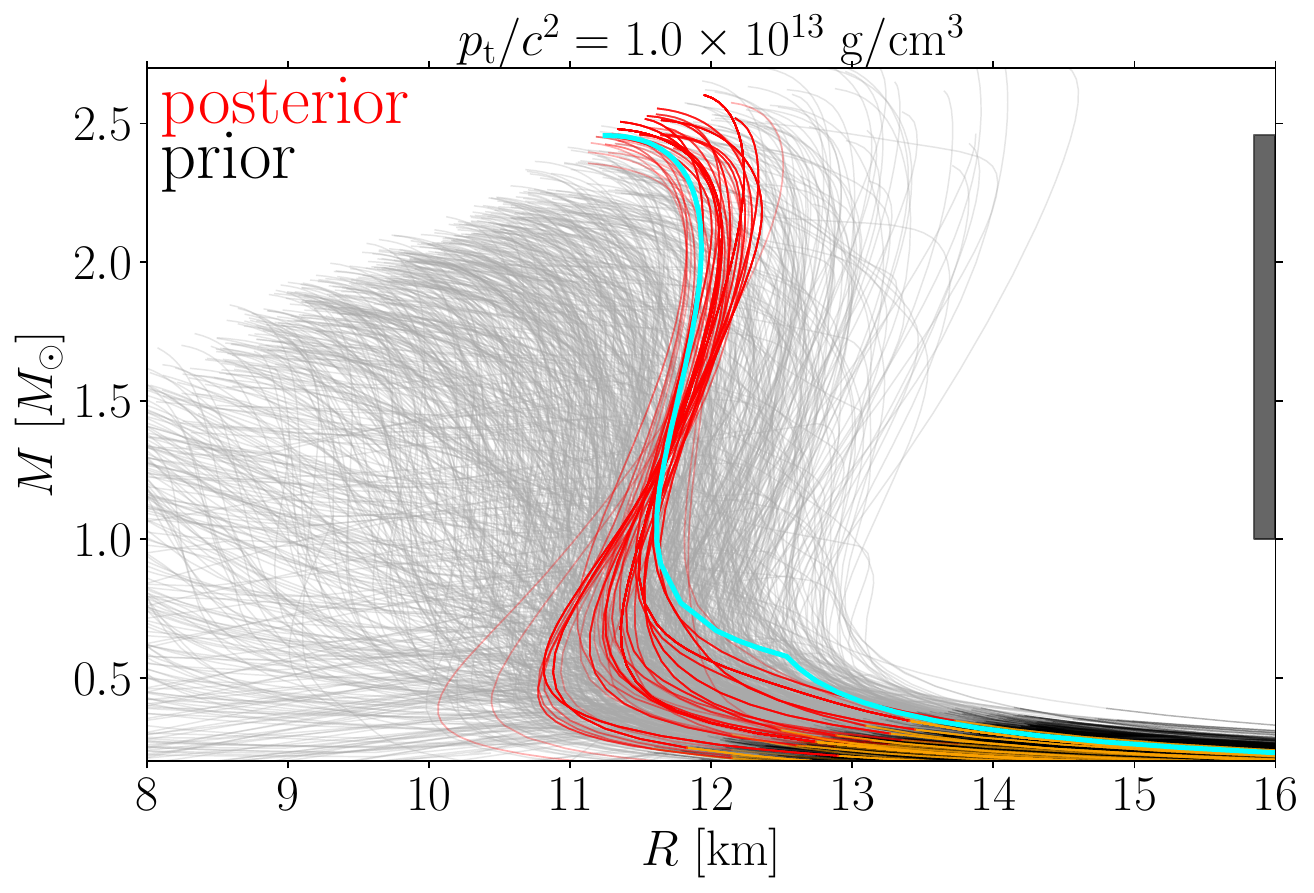}
        \includegraphics[width=0.49\linewidth]{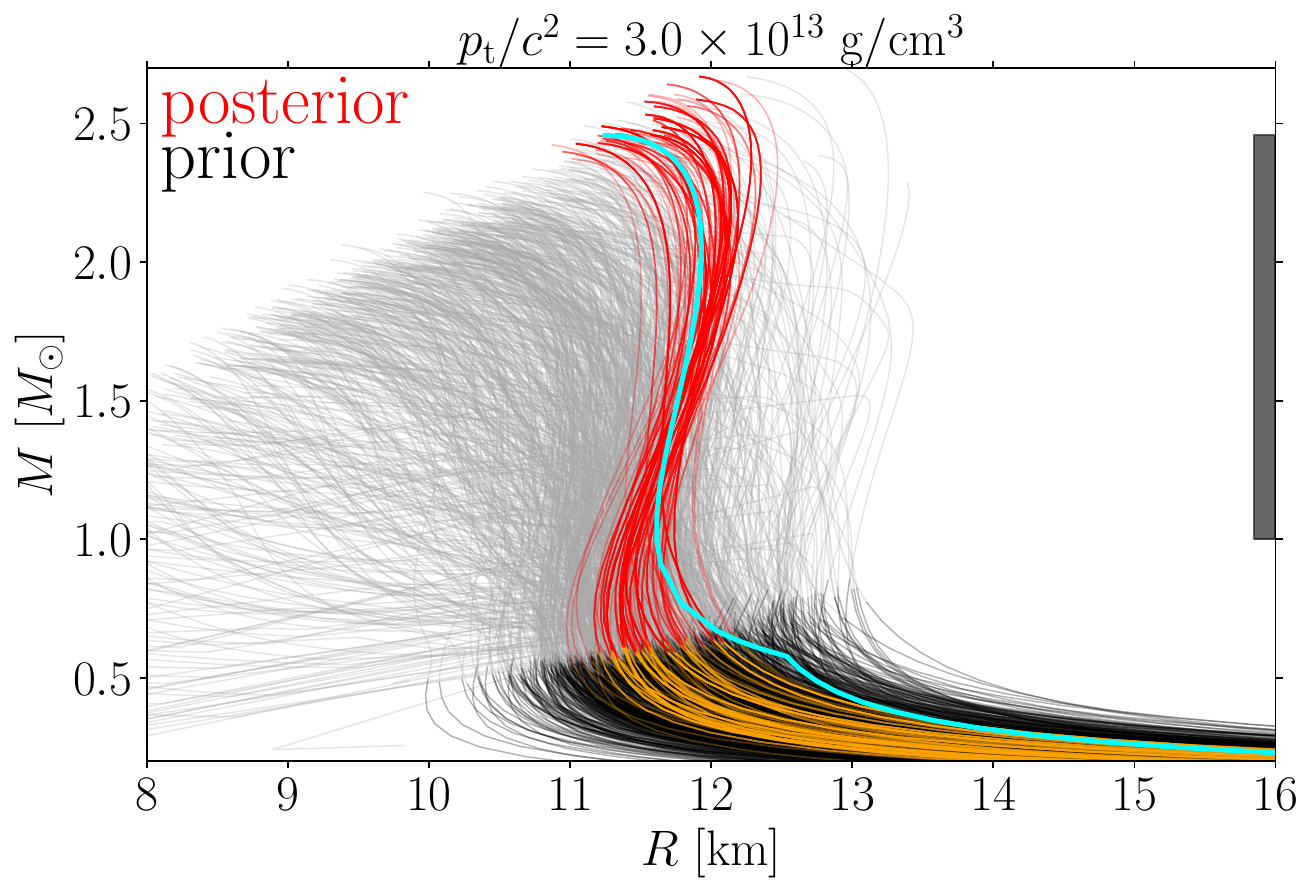}
        \includegraphics[width=0.49\linewidth]{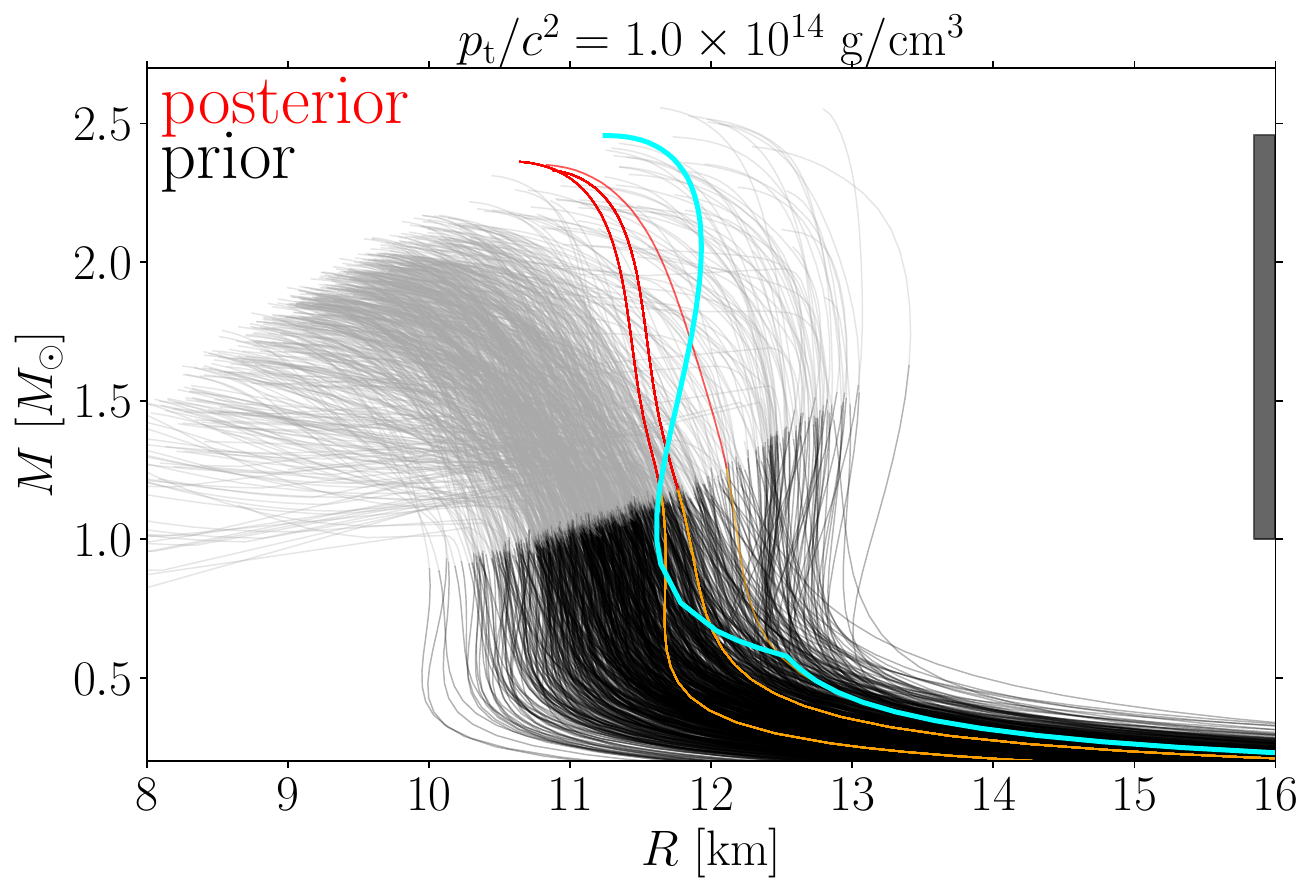}                

        \caption{Same as Fig.~\ref{fig:rmf-injected-posteriors}, but for a simulated EoS that undergoes a phase transition from an RMFT EoS to a constant speed of sound model, shown in cyan.  
        The RMFT description of the EoS holds only up to pressures slightly higher than the transition pressure ($\sim 1.8 \times 10^{13} \rm{g}/\rm{cm^3}$, which appears as a ``kink" in the injected EoS). We plot 200 fair draws from the posterior and 1000 from the prior.  We sample with replacement; if the same EoS from the posterior is sampled more than once (which happens generically if the posterior has few effective samples), the opacity of that EoS is proportional to the multiplicity of that sample. The gray bar has the same interpretation as in Fig.~\ref{fig:rmf-injected-posteriors}, except in this case the simulation EoS maximum mass is higher, so the bar extends to higher mass.    
        }
        \label{fig:rmf-pt-injected-posterior}
\end{figure*}

We now switch to the case of a simulated EoS which is not derived from an RMFT model, and repeat the analysis of Sec.~\ref{sec:rmf-recovery} using the phase-transition EoS described in Sec.~\ref{sec:description-of-inference}.
We display fair draws from the prior and posterior distributions on the mass-radius relation in Fig.~\ref{fig:rmf-pt-injected-posterior}.  
The priors are the same as in Fig.~\ref{fig:rmf-injected-posteriors}.
We find that as long as the transition pressure is sufficiently low, near or below the pressure at which the simulation EoS undergoes a phase transition ($p_{\rm t}/c^2 \sim 2.3\times 10^{13}\, \rm{g}/\rm{cm^3}$), the EoS is recovered effectively. 
By this, we mean that the mass-radius curves sampled from the posterior are reflective of the simulation EoS. 
For astrophysical neutron stars $M \in (\sim 1, M_{\rm TOV})$, the radius is near the center of the distribution.  

The phase-transition EoS represents a point of low prior probability in all EoS priors. 
This leads to poorer sampling resolution than in the case of recovering an EoS without a phase transition, particularly for the maximally agnostic cases.\footnote{See also Appendix C of Ref.~\cite{Essick:2023fso}.}
For example, for the case of $p_{\rm t}= 10^{13}\,\rm{g}/\rm{cm}^3$ and  $3\times 10^{13}\,\rm{g}/\rm{cm}^3$ we have $\sim 11$ and $\sim 33$ effective samples,\footnote{We define the number of effective samples $n_{\rm eff} \equiv 
(\sum_i w_i)^2/(\sum_iw_i^2)$.  Because the prior is uniform, this is closely related to the Kullback-Liebler (KL) divergence~\cite{Kullback1951}, $\mathrm{KL}(p||q) \equiv \sum_x p(x)\ln(p(x)/q(x))$.  The KL divergence from prior to posterior for the PT-injection case recovered with the model transitioning at $1\times 10^{12}\,\rm{g}/\rm{cm^3}$ is $\sim 7.5$, the analogous KL value for the case with the RMFT injection is 3.7.  This is another way of stating that the injected PT EoS is a poor representative of the entire EoS prior, since the posterior centered on it is far from the EoS prior. } respectively.    
Despite this, macroscopic observables are well recovered.
In the cases above, $R_{1.4} = 11.76^{+0.09}_{-0.11}\,\rm{km}$, and $R_{1.4} = 11.73^{+0.12}_{-0.14}\,\rm{km}$ for $p_{\rm t}= 10^{13}\,\rm{g}/\rm{cm}^3$ and  $3\times 10^{13}\,\rm{g}/\rm{cm}^3$, respectively (quoted at $90\%$ credibility), consistent with the simulated value $R_{1.4} =  11.72~\rm{km}$. 
For the case of a transition pressure $p/c^2 = 10^{14} \rm{g}/\rm{cm}^3$, however, the EoS is no longer recovered in that the posterior no longer reflects the simulation EoS within uncertainties.
The maximum mass of all EoSs in the posterior is below the simulation EoS maximum mass.
Furthermore, the posterior has only $\sim 2$ effective samples.

\begin{figure}
    \centering
            \includegraphics[width=0.49\textwidth]{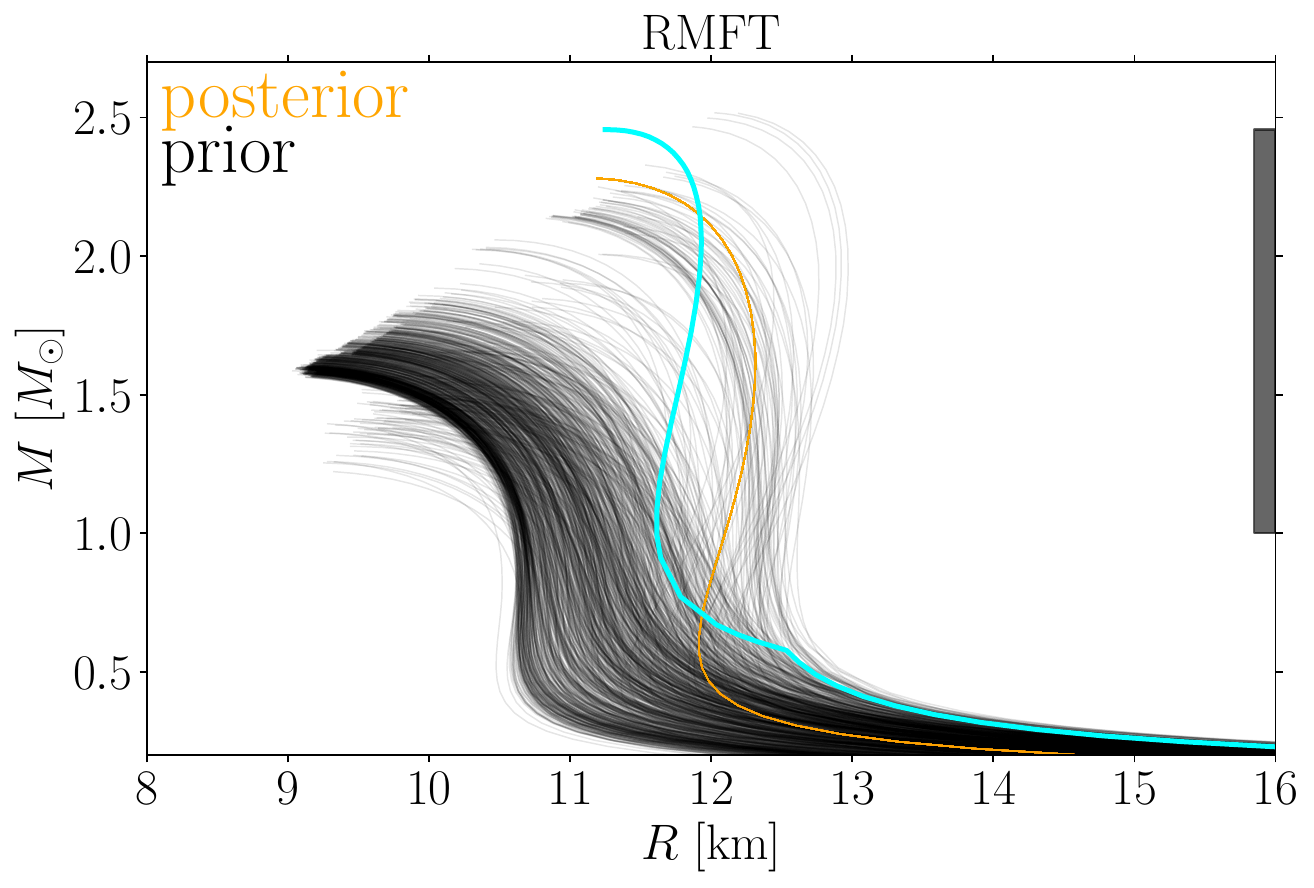}
    \caption{Same as Fig.~\ref{fig:rmf-pt-injected-posterior}, but for the RMFT prior distribution (same as Fig.~\ref{fig:rmf-inj-rmf-posterior}).  The RMFT is unable to recover the non-RMFT simulation EoS.  There is only one posterior EoS displayed as the posterior contains only a single effective sample. 
    }
    \label{fig:pt-inj-rmf-posterior}
\end{figure}

For comparison, we display the result of the identical inference using the RMFT EoS set as a prior in Fig.~\ref{fig:pt-inj-rmf-posterior}.  
The result is, in effect, the same as the result for the hybrid RMFT-informed model-agnostic posterior which transitions at $p/c^2 = 10^{14} \rm{g}/\rm{cm}^3$, as there are too few effective samples from the analysis to even form a reasonable estimate for the posterior.  
We therefore conclude that the RMFT EoS prior, like the hybrid prior transitioning at $10^{14}\,\rm{g}/\rm{cm}^3$, is effectively inconsistent with the simulated data.

\begin{figure}
    \centering
    \includegraphics[width=\linewidth]{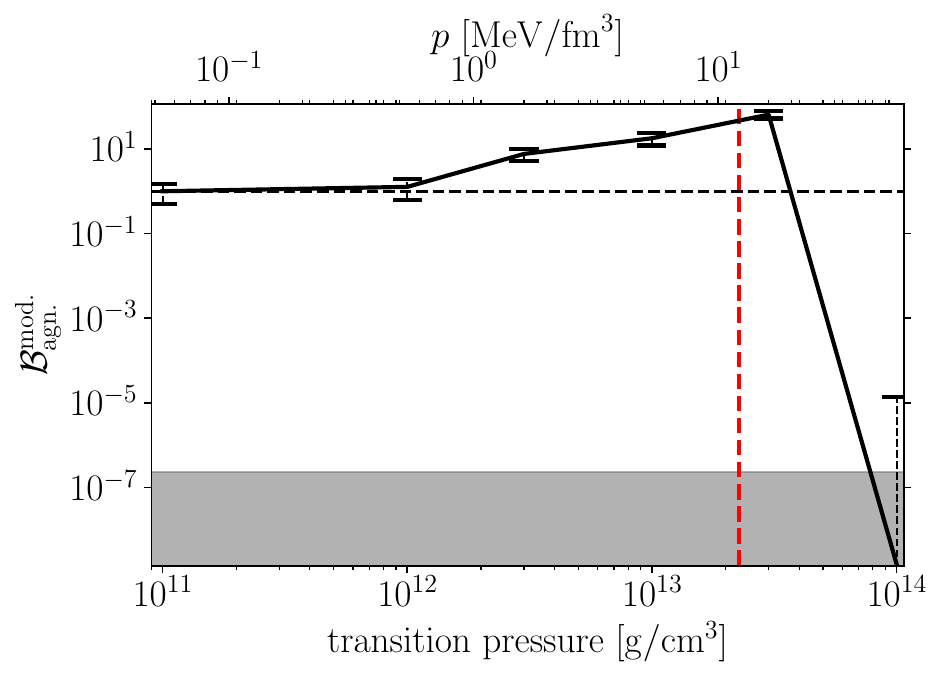}
    \caption{Same as Fig.~\ref{fig:rmf-bayes-factors}, but for a simulated EoS with a strong phase transition to a constant speed-of-sound near 1.5 times saturation density (red vertical line).
    }
    \label{fig:rmf-pt-bayes-factors}
\end{figure}

Figure~\ref{fig:rmf-pt-bayes-factors} shows the Bayes factor for each model with a different transition density. 
Most Bayes factors represent inconclusive evidence in favor of the RMFT description, with this evidence becoming stronger at higher transition pressures. 
This makes sense, as the low-density EoS is indeed well-described by the RMFT.  
For transition pressures that favor the RMFT-informed EoS over model-agnostic EoS, we find larger Bayes factors than the corresponding values in Fig.~\ref{fig:rmf-bayes-factors}.
This is due to two factors.  First, one particular simulated x-ray measurement is highly informative (phase transition injection ``X-ray 1", in Tab.~\ref{tab:ns_observations_pt}), and more precise than the most informative RMFT EoS injection by about a factor of 40\%.  
This, however, is not sufficient to explain the very large difference in evidence. 
Removing this event leaves maximum Bayes factors of order $\sim 10$, still larger than those of Fig.~\ref{fig:rmf-bayes-factors}.

The second cause of the large Bayes factors is the unusual location of the phase-transition EoS relative to all of the EoS priors used.  Because the phase-transition EoS is ``unusual" among both model-agnostic and RMFT EoSs, it lies at the margins of all of the prior distributions on the EoS.  
At these margins, small changes to the prior will lead to large changes in the (log) likelihood of the bulk of EoSs, analogous to how small changes in the z-score dramatically change the (log) probability of a Gaussian far from the mean.  
These large changes in evidence lead directly to large Bayes factors.  
These Bayes factors are indicative that the RMFT is the correct underlying description of the low-density EoS, but (comparing to Fig.~\ref{fig:rmf-bayes-factors}), the size of the Bayes factors are due to the details of the prior construction and the location of simulation EoS.

For transition pressure $p_{\rm{t}}/c^2 = 10^{14}\,\rm{g}/\rm{cm}^3$, we recover strong evidence against an RMFT description. 
While uncertainties are large, the Bayes factor for the $p_{\rm{t}}/c^2 = 10^{14}\,\rm{g}/\rm{cm}^3$ model relative to the agnostic model is less than $10^{-4}$ at 90\% credibility.
This quantifies the statement that the RMFT-informed EoS prior is unable to produce candidate EoSs that closely mimic the simulation EoS, leading to few (or no) EoSs that are consistent with all astrophysical observations.
Further, this confirms that if there is a strong phase transition, then we can identify the associated breakdown of the RMFT using our procedure given sufficient astrophysical data.

\begin{figure}
    \centering
    \includegraphics[width=0.49\textwidth]{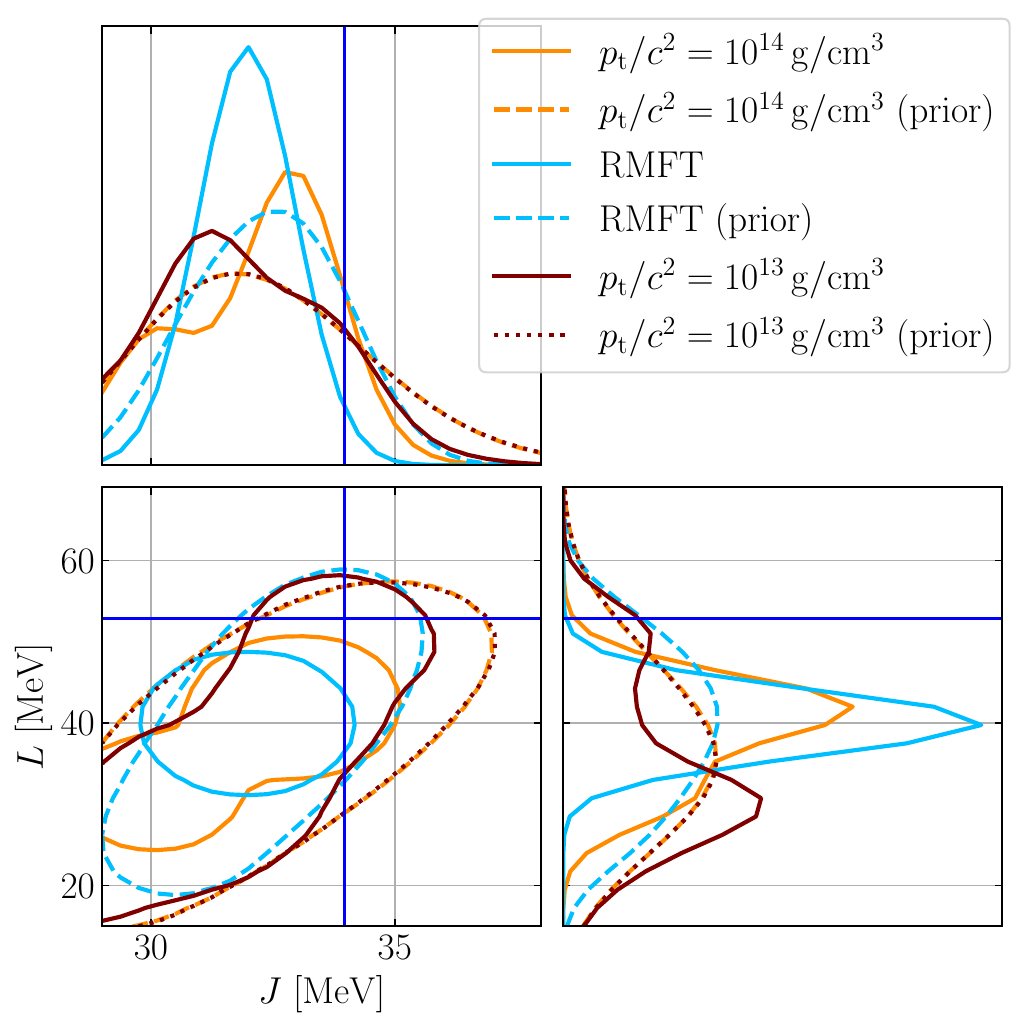}
    \caption{Same as Fig.~\ref{fig:rmf-symmetry-params} but with a simulation EoS which transitions to a constant speed of sound EoS.
    }
    \label{fig:rmf-pt-symmetry-params}
\end{figure}

Finally, we show the inferred symmetry parameters in Fig.~\ref{fig:rmf-pt-symmetry-params} for different transition pressures and when using the set of RMFT EoSs itself.  
Because we use the same RMFT parameters for the low-density EoS, the simulation EoS's symmetry parameters are the same as in Fig.~\ref{fig:rmf-symmetry-params}.
Moreover, since we use the same EoS priors, the marginal priors on $J$ and $L$ are consistent with the simulation EoS's values.
Nonetheless, using the RMFT EoS set or the hybrid prior with $p_{\rm{t}}/c^2 \sim 10^{14}\,\rm{g}/\rm{cm}^3$ does not recover the correct symmetry parameters.
The correct symmetry parameters are recovered only for the GP EoS distribution that transitions at a lower pressure. 
This is because at low transition pressures, the marginal distribution on the symmetry parameters is effectively unchanged by the inclusion of astrophysical observations, indicating that astrophysical observations carry little information about the symmetry parameters under these models~\cite{Essick:2021kjb, Essick:2021ezp}.
When the RMFT description is trusted to higher pressures, however, the inclusion of astrophysical data renders the marginal distribution on the symmetry parameters inconsistent with the simulation EoS. 
This is reasonable, as the astrophysical properties of neutron stars with the simulation EoS are primarily determined by the ``quark matter" (constant speed of sound EoS), not the low-density RMFT (hadronic) EoS.
As such, inferring the properties of the hadronic EoS assuming it holds up to high pressures will lead to a bias.

The cases of two possible EoSs with and without a phase transition represent two possible examples of astrophysically realistic EoSs.  
If the astrophysical EoS has a stronger phase transition than the one we simulated, we expect to find evidence with fewer data. 
We have chosen the parameters of the quark matter EoS in such a way that no unstable or twin branches arise, but a connected stable hybrid branch (case (c) in Fig.\,2 of Ref.~\cite{Alford:2015gna}). 
This presents the greatest challenge, as twin or unstable branches would be easier to distinguish. 
If, on the other hand, there is a weaker phase transition, much more precise observations may be necessary to achieve similar evidence.  
To give a sense of the sensitivity of our analysis to the choice of the EoS phase transition parameters, we perform the same analysis with an EoS with a weaker transition in Appendix.~\ref{sec:additional-analysis}.  
We find that even in cases where astrophysical observations are unable to decisively identify a breakdown scale, they may still imply a posterior inconsistent with nuclear physics, in which case our approach will provide hints about the possible sources of the breakdown of the model.

\section{Constraints on RMFT-breakdown with current astrophysical data}
\label{sec:astro-analysis}

\begin{figure*}
 \centering
 \includegraphics[width=0.49\linewidth]{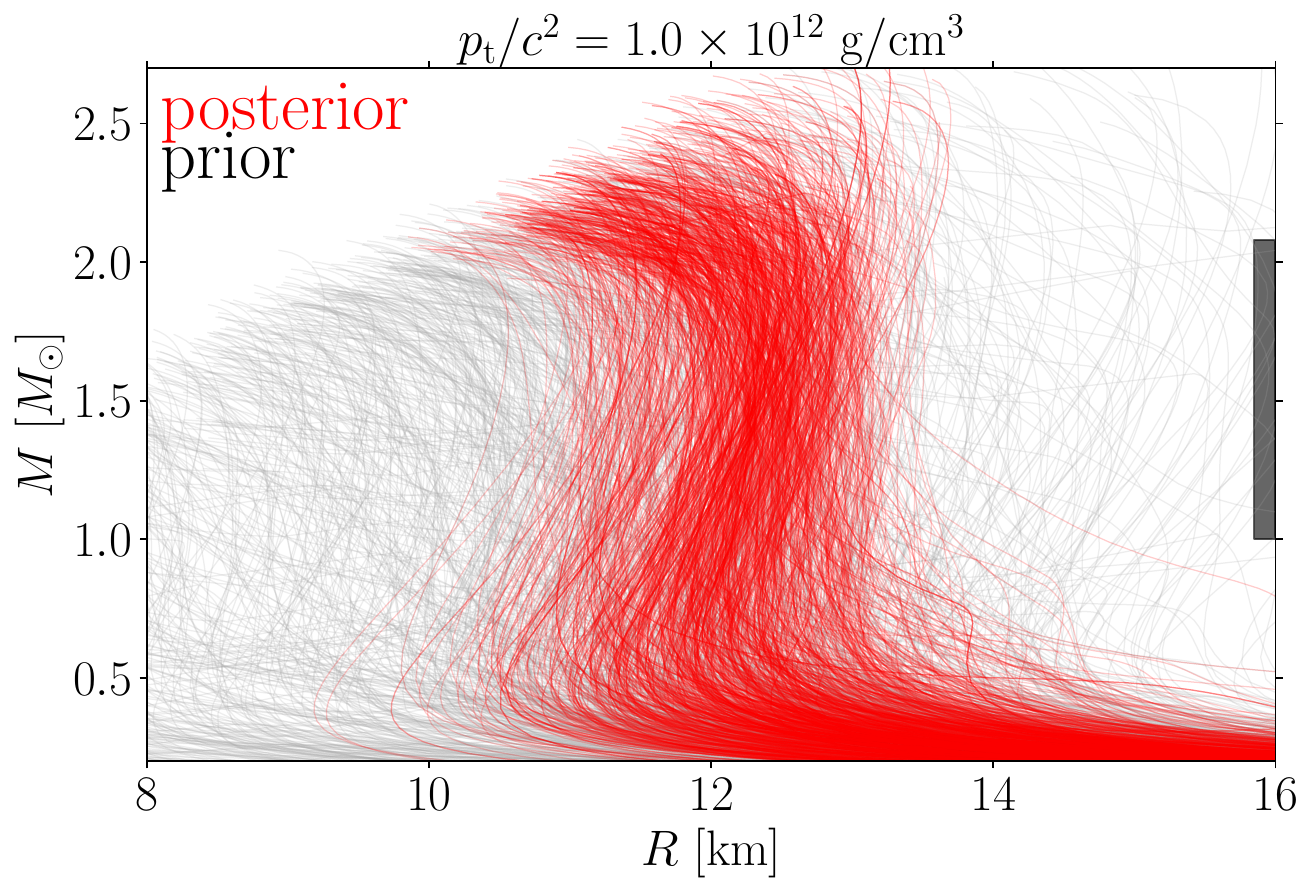}
   \includegraphics[width=0.49\linewidth]{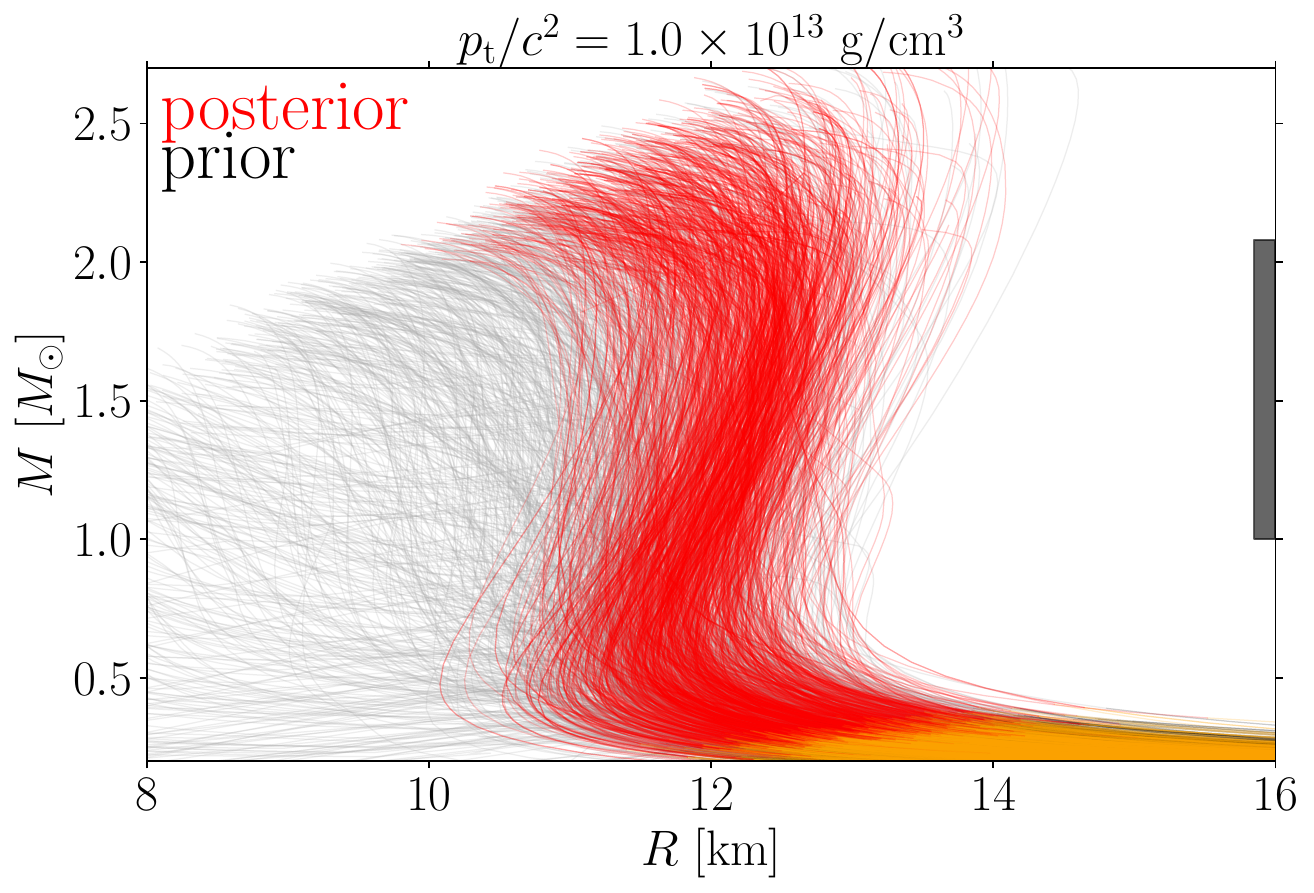}
    \includegraphics[width=0.49\linewidth]{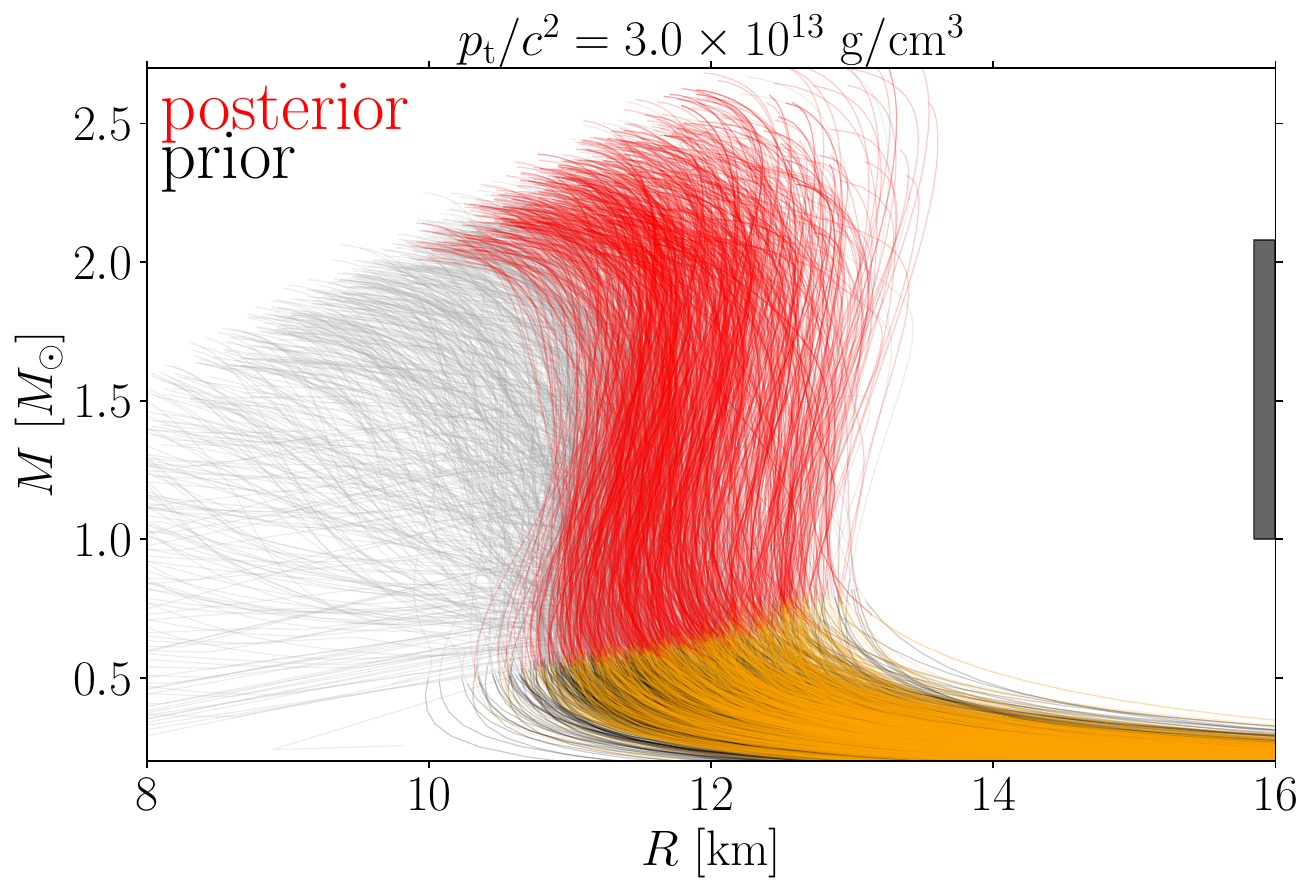}
    \includegraphics[width=0.49\linewidth]{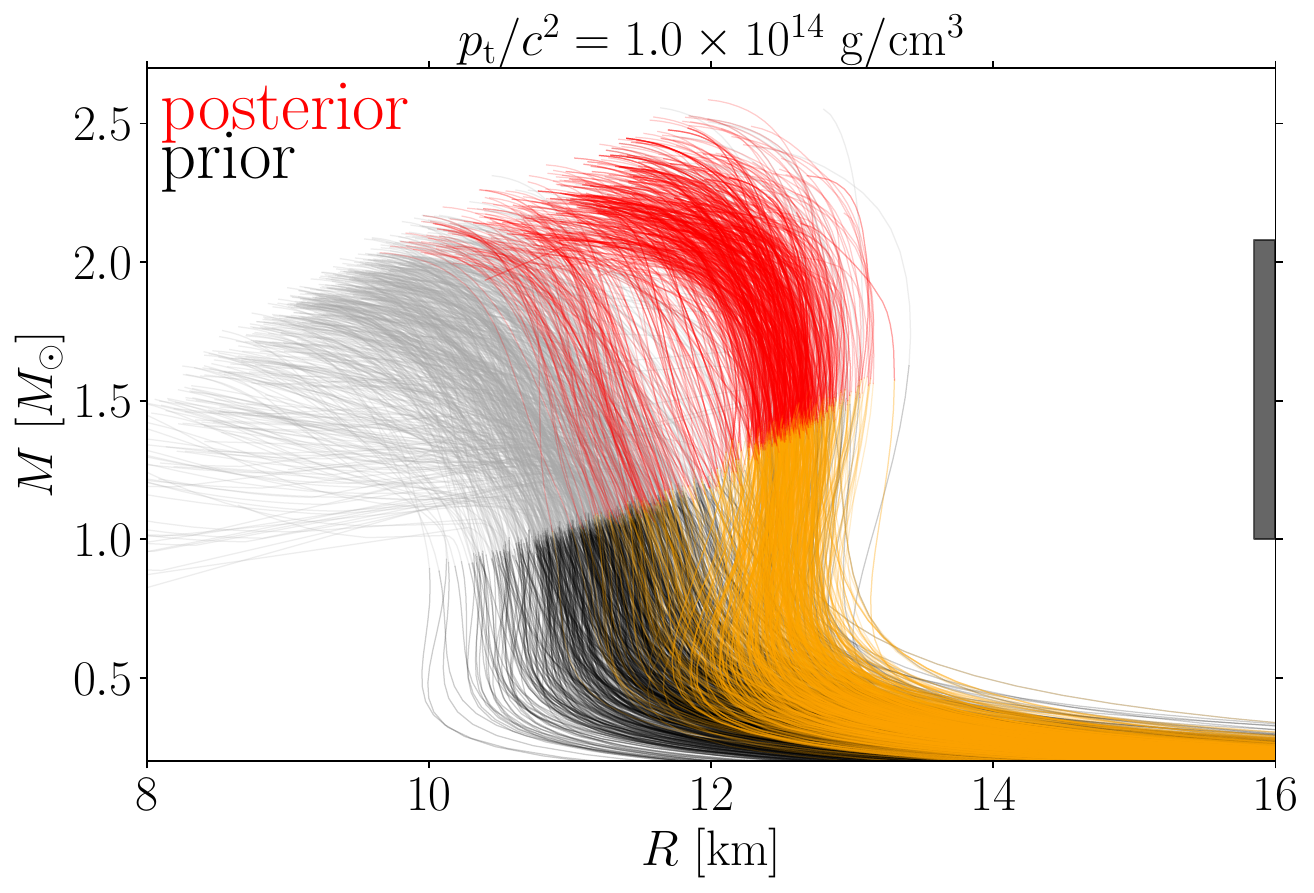}
    \caption{Same as Fig.~\ref{fig:rmf-injected-posteriors}, but with real astrophysical data instead of simulated data. 
    }
    \label{fig:astro-posterior}
\end{figure*}

We finally turn to current astronomical data: radio timing observations of J0348+0432~\cite{Antoniadis:2013pzd}, NICER x-ray pulse profile observations of J0030+0451~\cite{Miller:2019cac}, J0740+6620~\cite{Miller:2021qha}, and J0437-4715~\cite{Choudhury:2024xbk}, and gravitational-wave events GW170817~\cite{gw170817} and GW190425~\cite{gw190425}.   
We display fair draws from the prior and posterior for various choices of the transition pressure in Fig.~\ref{fig:astro-posterior}. 
The posteriors are wider than in the simulation studies, since the current data are not as constraining.
We find larger uncertainties at lower transition pressures, consistent with expectations that the RMFT-informed distribution imposes tighter constraints \emph{a priori} than the model-agnostic GP.
This, however, depends on mass, as $R_{1.4}$ is less variable between models than $R_{1.0}$, for example.  

\begin{figure*}
    \centering
    \includegraphics[width=0.9\linewidth]{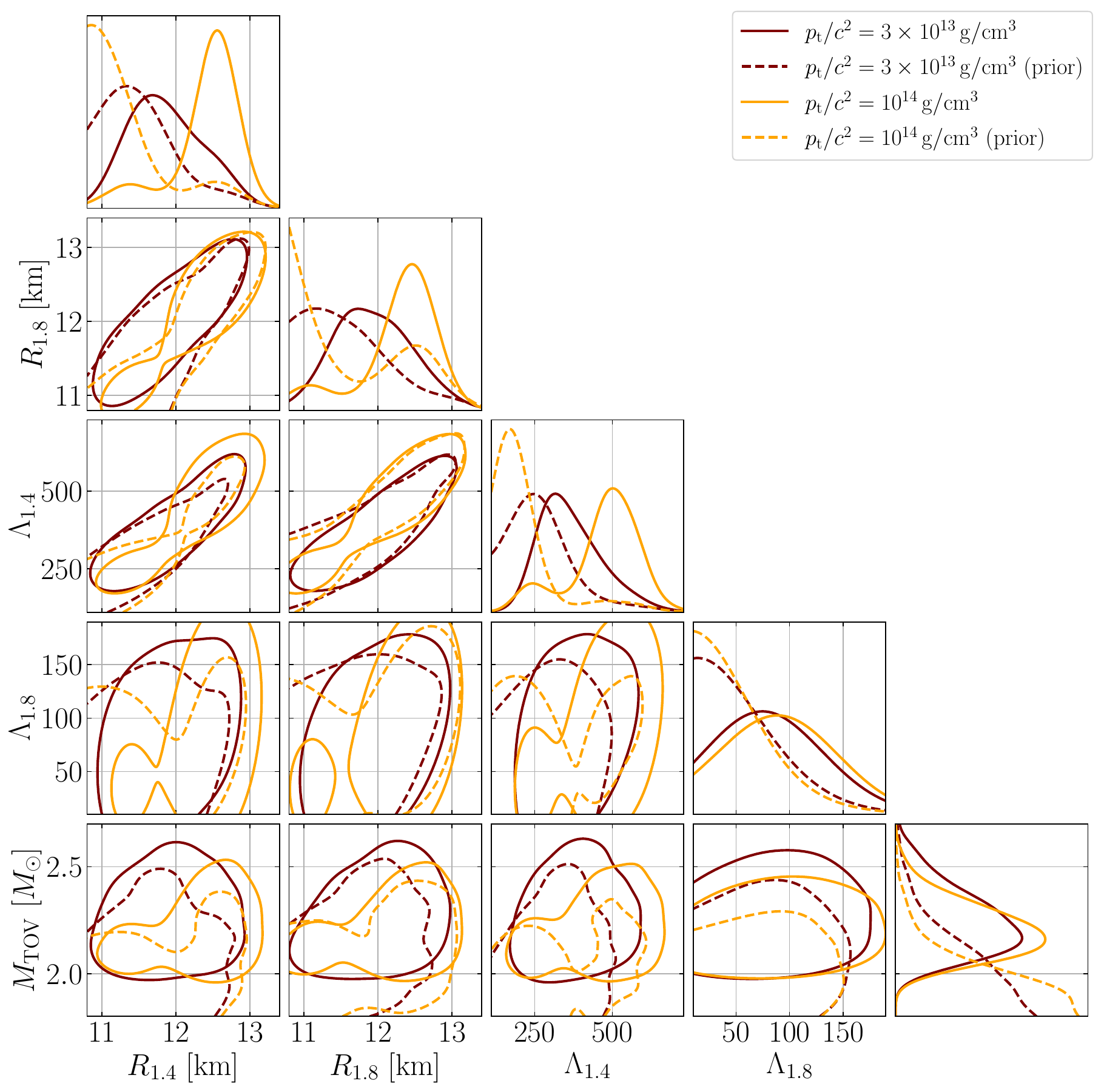}
    \caption{Prior (dashed) and posterior (solid) distributions for select parameters when transitioning from the RMFT-conditioned to the model-agnostic EoS prior at $3\times 10^{13}\, \mathrm{g}/\mathrm{cm}^3$ (maroon) and $10^{14}\, \mathrm{g}/\mathrm{cm}^3$ (orange).  Lines mark $90\%$ credible regions.  We show: the radius ($R_{1.4},\, R_{1.8}$) and dimensionless tidal deformability ($\Lambda_{1.4}, \, \Lambda_{1.8}$) of 1.4 and 1.8$\,M_{\odot}$ neutron stars respectively and the maximum TOV mass ($M_{\rm TOV}$) of a neutron star.
    } 
    \label{fig:astro-params-corner}
\end{figure*}

In Fig.~\ref{fig:astro-params-corner} we show the prior and posterior distributions on various astrophysical quantities under different assumptions on the transition density.  
Trusting the RMFT to higher pressures favors a larger value of $R_{1.4}$ and $\Lambda_{1.4}$.  
For example, $R_{1.4} = 11.80^{+0.92}_{-0.64}\,\rm{km}$ for the model that transitions at $3 \times 10^{13}\, \rm{g}/\rm{cm}^3$, whereas $R_{1.4} = 12.51^{+0.34}_{-1.21}\,$km for the model transitioning at $10^{14}\,\rm{g}/\rm{cm}^3$.    
This difference is a consequence of stronger correlations between lower and higher densities, which translates to stronger correlations between ${R_{1.4}}$ and $M_{\rm TOV}$ under the RMFT.  
Under the assumption of a transition density of $10^{14}\, \rm{g}/\rm{cm}^3$, values of $M_{\rm TOV}$ greater than $\sim 2.2\,M_\odot$ are inconsistent with small neutron star radii (e.g. $R_{1.4} \lesssim 12\,\rm{km}  $) \emph{a priori}.  
This is not true for a lower transition pressure. 
Therefore, similar $\MTOV$ posteriors result in different $R_{1.4}$ posteriors due to these correlations between high and low densities.   

On the other hand, $\Lambda_{1.8}$ is inferred to be very similar between the two models.
This is because the tidal deformability is a strongly decreasing ($\Lambda \sim m^{-6}$~\cite{Hinderer:2009ca, Zhao:2018nyf}) function of the mass.
Therefore, EoS differences are much suppressed in $\Lambda_{1.8}$.
This convergence does not happen for $R_{1.8}$, however, which shows qualitatively the same picture as $R_{1.4}$.   
Therefore, even at constant fractional uncertainty in tidal parameters (which requires much louder gravitational wave signals), we expect neutron star mergers will be much better probes of RMFT breakdown at low masses than at high masses. 
Radius constraints, on the other hand, are similarly constraining at all masses.

\begin{figure}
    \centering
    \includegraphics[width=\linewidth]{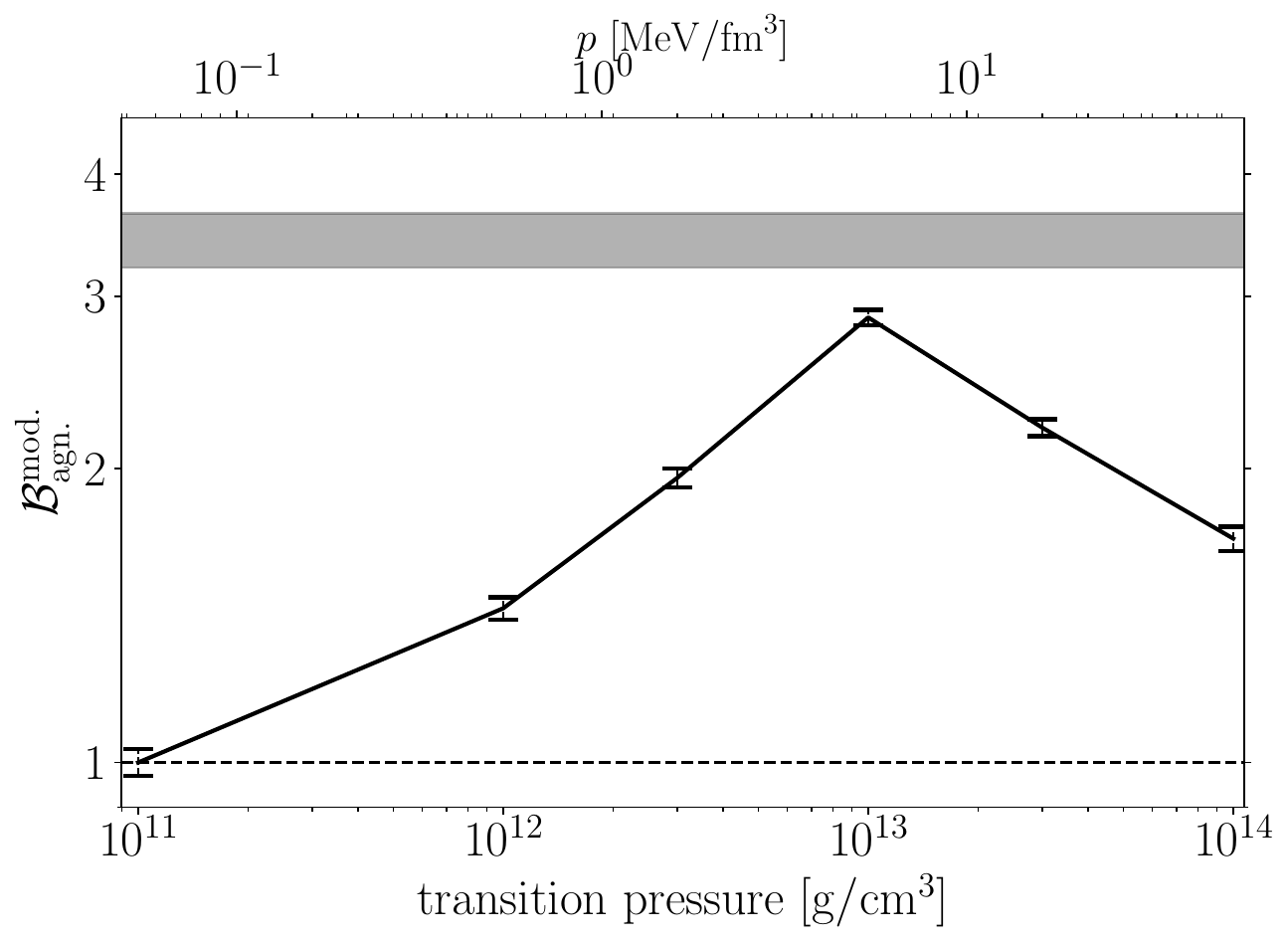}
    \caption{Same as Fig.~\ref{fig:rmf-bayes-factors}, but using astrophysical observations rather than simulated data.  We find no strong preference for any transition pressure.}
    \label{fig:astro-bayes-factor}
\end{figure}

 In Fig.~\ref{fig:astro-bayes-factor}, we show the Bayes factors of each transition pressure compared to the lowest transition pressure. 
 We generally find a preference for higher transition pressure with a mild decrease in evidence at the highest transition pressure, although the Bayes factors are all 2-3. 
Therefore, we do not find evidence against an RMFT description of the astrophysical EoS up to  $p/c^2 \sim 10^{14}$\,$\rm{g}/\rm{cm}^3$ using current astrophysical data.  
We additionally do not find strong evidence for the RMFT, though as discussed in Sec.~\ref{sec:rmf-recovery}, we do not necessarily expect such evidence if RMFT is the correct underlying model.

\begin{figure}
    \centering
    \includegraphics[width=0.5\textwidth]{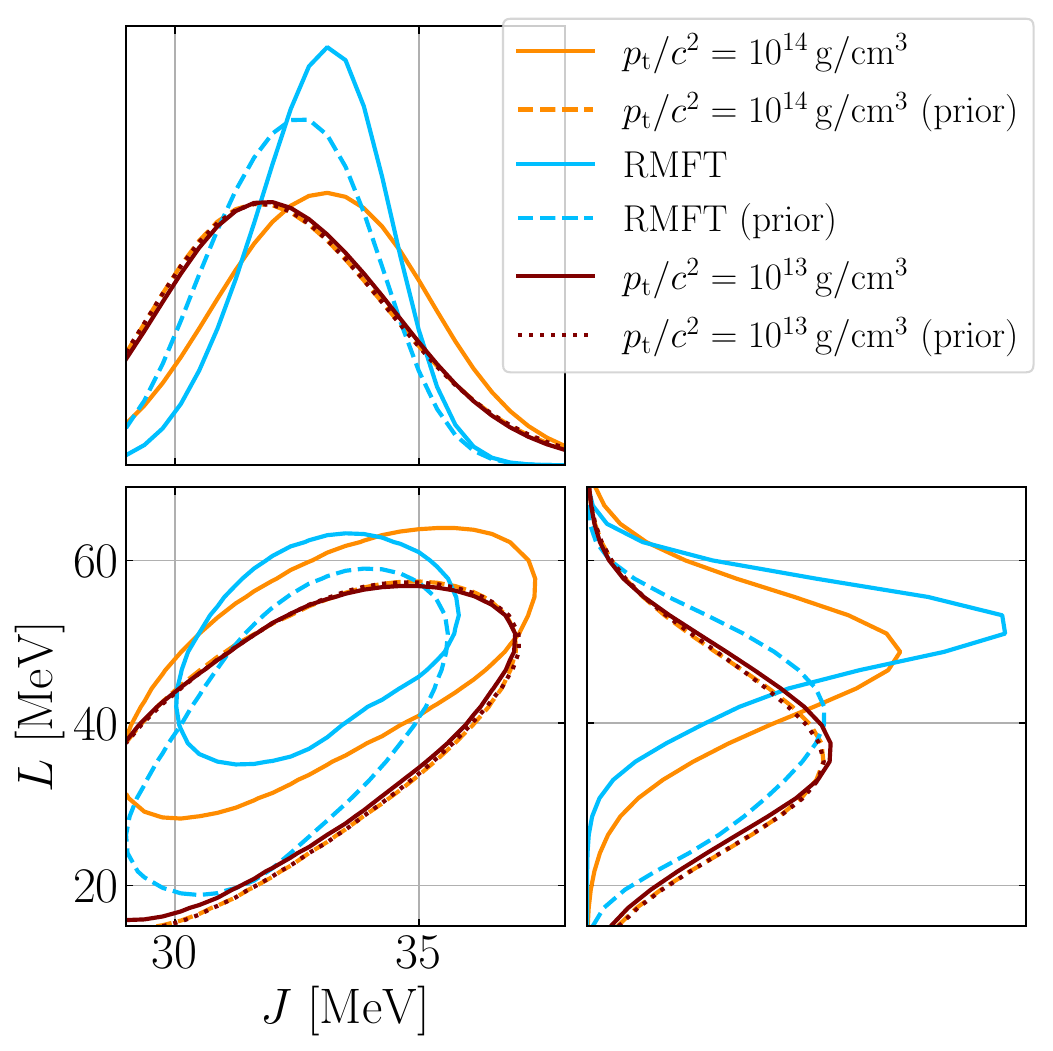}
    \caption{The inferred symmetry parameters using astrophysical observations, same as Fig.~\ref{fig:rmf-symmetry-params}. The prior for the model which transitions at $ 10^{13}\, \rm{g}/\rm{cm}^3$ (maroon dashed) is essentially the same as the posterior and the prior which transitions at $10^{14}\, \rm{g}/\rm{cm}^3$  (same as in Fig.~\ref{fig:rmf-symmetry-params}), and therefore we mark it with a dotted line to increase contrast.
    }
    \label{fig:astro-symmetry-params}
\end{figure}

The symmetry energy and its slope at saturation are shown in Fig.~\ref{fig:astro-symmetry-params} for the RMFT EoS set and hybridized priors transitioning at $p_{\rm t}/c^2 = 10^{13}\, \rm{g}/\rm{cm}^3$ and $10^{14}\, \rm{g}/\rm{cm}^3$. 
The posteriors are largely consistent with each other, though for higher transition pressures, and for the RMFT EoS set itself, the values of both $L$ and $J$ are better constrained.  
We find, for example, that $L = 37.03^{+16.37}_{-14.42}\,$MeV when we transition at $10^{13}\,\rm{g}/\rm{cm}^3$, and $L = 47.87^{+11.15}_{-14.04}\,$MeV when transitioning at $10^{14}\,\rm{g}/\rm{cm}^3$.
The recovered value of $L$ is both larger and better constrained when we trust the RMFT up to higher densities, likely due to correlations between the large inferred maximum mass of neutron stars and the stiffness of matter near saturation under these models.  
Trusting the RMFT 
up to only $p_{\rm t}/c^2 = 10^{13}\,\rm{g}/\rm{cm}^3$, which corresponds to a density $\sim 1.5\, \nsat$, we find no meaningful constraints on the symmetry parameters relative to the priors.
This is an indication that under the model-agnostic prior, the properties of matter near saturation are not strongly correlated with the matter in the cores of astrophysical neutron stars~\cite{Essick:2021kjb, Essick:2021ezp}.

The recovered values are consistent with existing results on the symmetry parameters, although they represent matter which is marginally softer near saturation than other studies.
Reference~\cite{Drischler:2020hwi}, using constraints from \chiEFT{} at N$^3$LO, finds $J = 31.7 \pm 1.1\, \rm{MeV}$ and $L = 59.8 \pm 4.1\, \rm{MeV}$ at $1\sigma$ uncertainty. Additionally, per Fig.~2 of Ref.~\cite{Drischler:2020hwi} (based partly on Refs.~\cite{Lattimer:2012xj,Lattimer:2014sga}) this range is also found to be largely consistent with a broad range of theoretical, experimental, and observational constraints of the nuclear symmetry energy. 
Our results are also consistent with values implied by PREX-II~\cite{Adhikari:2021phr}, though we do not consider that data here.
For instance, Ref.~\cite{Reed:2021nqk} used PREX-II results to infer $L=106 \pm 61\, \rm{MeV}$ at $90\%$ credibility.
Combining PREX-II and low-density $\chi$EFT information in the GP framework yields $L=53^{+14}_{-15}\,\rm{MeV}$~\cite{Essick:2021kjb, Essick:2021ezp}. Using a less flexible EoS model, though with a wider range of astrophysical and nuclear data including PREX-II, Ref.~\cite{Biswas:2024hja} found $L=54^{+10}_{-10}\,\rm{MeV}$ at $90\%$.   

\begin{figure}
    \centering \includegraphics[width=0.5\textwidth]{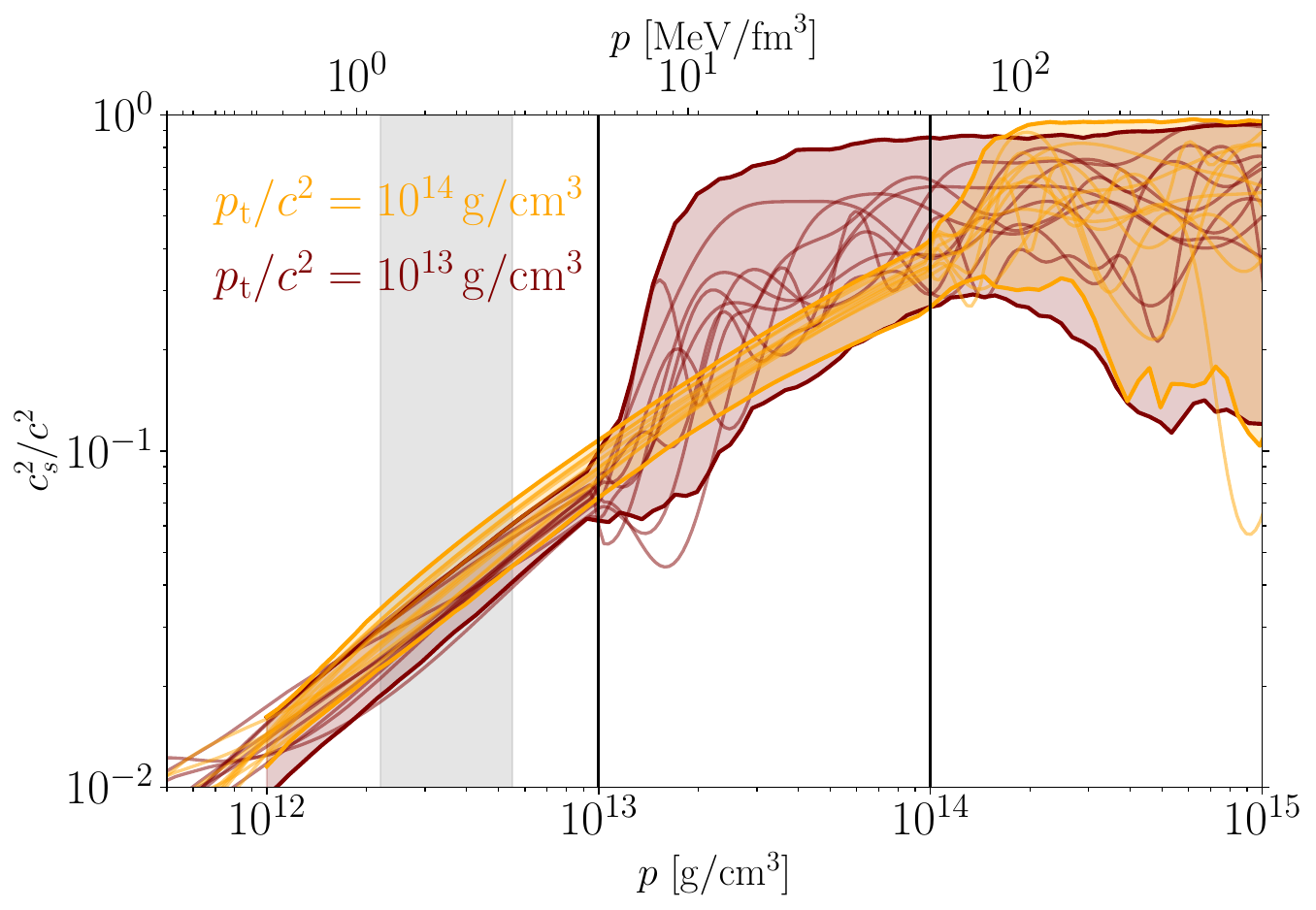}
    \caption{Speed-of-sound squared 90\% credible intervals using current astrophysical data, along with fair draws for the prior transitioning to agnostic at $10^{13}\,\rm{g}/\rm{cm}^3$ and $10^{14}\, \rm{g}/\rm{cm}^3$. Transition pressures are marked by black vertical lines. A gray bar shows the approximate range of pressures at saturation density for the RMFT EoSs.
    }
    \label{fig:astro-rho-vs-cs2}
\end{figure}

Even though both the GPs which transition at $10^{13}\,\rm{g}/\rm{cm}^3$ and $10^{14}\,\rm{g}/\rm{cm}^3$ are described by the same RMFT framework near saturation density, we recover different posteriors on the EoS in this region.  
This is because correlations with the well-constrained high-density EoS are different under the two choices of model.  
To further show this, in Fig.~\ref{fig:astro-rho-vs-cs2} we display the inferred sound speed as a function of density under different choices of the transition pressure.  
The sound speed at saturation density is markedly different for the two posteriors, even though both of the distributions are closely emulating the underlying RMFT at that density and therefore have very similar prior distributions on $c_s^2$.  This happens because higher-density observations can more strongly inform low-density physics when the RMFT description is trusted up to higher densities.

\section{Discussion}
\label{sec:discussion}
 

We have constructed hybrid models of the dense-matter EoS informed by an RMFT at low pressures and model-agnostic above a certain transition pressure.
We compare multiple models with different transition pressures to astrophysical data, and probe the scale up to which a specific RMFT EoS is able to describe the dense-matter EoS.
We first explore how inference behaves under simulated data.
If the true EoS is consistent with the RMFT description, evidence will remain inconclusive, unable to rule out the RMFT at any scale.
This behavior is expected from prior volume arguments.
If the true EoS is inconsistent with any RMFT EoS, for example due to a phase transition,
we recover strong evidence (Bayes factors $\gtrsim 10^4$) against models that are informed by the RMFT beyond the pressure of the phase transition.  However, the strength of the evidence is sensitive to the parameters of the phase transition; in general, stronger transitions lead to stronger evidence of RMFT-breakdown with fewer observations.
We further demonstrate that a breakdown can be identified with sufficiently many gravitational-wave observations in App.~\ref{sec:simulated-inference-gws}, within expectations of upcoming observing campaigns.

Applying our models to real astrophysical observations, we find no evidence against RMFT EoSs up to arbitrarily high densities. When we assume no breakdown of the RMFT model, our inferred EoS distribution is qualitatively consistent with existing constraints on RMFT EoSs using astrophysical data~\cite{Traversi:2020aaa, Malik:2022zol, Huang:2023grj, Zhou:2023hzu, Xia:2024nwa}.
Nonetheless, the inferred macroscopic neutron star properties differ depending on the pressure up to which the RMFT is trusted.
This behavior elucidates how the breakdown pressure can be constrained with future data.  
For example, the inferred radius of a $1.4~M_{\odot}$ star depends on what pressure the RMFT is trusted up to.
Conditioning up to a pressure of $1 \times 10^{14}\,\rm{g}/\rm{cm}^3$, we find $12 \lesssim
 R \lesssim
 13\,\rm{km}$ at $\sim 90\%$ credibility, while conditioning up to only $3\times 10^{13}~\rm{g}/\rm{cm}^3 $, yields $11 - 12.5\,\rm{km}$. 
This suggests that future constraints should be more informative about the breakdown scale if they reach $\sim 0.5\,\rm{km}$ precision and $R_{1.4}\sim 11.5\,\rm{km}$.
Finally, as a note of caution, it is possible that  biases emerge before evidence against a model appears. Therefore it is still important to consider other models of the EoS, including fully agnostic analyses. From this point of view, we see this work as complementary to existing studies which search for the breakdown of explicit models of low-density matter using astrophysical observations~\cite{Biswas:2020puz, Wouters:2025zju}.

If an RMFT-informed analysis is deemed inconsistent with astrophysical data, then the RMFT EoSs have broken down.
Possible reasons and remedies are listed in the introduction. 
Our method not only yields evidence for the breakdown, but can also quantify the breakdown pressure. 
For example, if the breakdown occurs at ${\sim}2-3\,n_0$, then it is possibly due to hyperon degrees of freedom \cite{Tolos:2020aln}. 
Alternatively, if the breakdown happens at ${\sim}4-6\,n_0$, then it is likely that the treatment of nucleons as point-like particles is insufficient as the average distance of the nucleons becomes smaller than their expected size~\cite{Weber:2004kj}.  
More generally, understanding the regime where the RMFT has broken down can be challenging and comparing multiple models 
will likely be necessary.

Further observables, such as the post-merger gravitational wave signal~\cite{Shibata:2005xz,Radice:2013hxh, bauswein:15, Breschi:2022xnc, Dietrich:2020eud}, an electromagnetic counterpart~\cite{Dietrich:2020efo, ZhouChen2019, CapanoTews2019, Coughlin:2018fis}, or the lifetime of the remnant~\cite{Margalit:2017dij, Shibata:2019ctb}, may also indicate that the breakdown is due to a strong phase transition.
In that case, a range of approaches, both modeled and model-agnostic, can be used to constrain its properties~\cite{Pang:2020ilf, Chatziioannou:2019yko, Annala:2019puf, Essick:2023fso, Mroczek:2023zxo}.  
We do not find evidence for a breakdown, and thus a strong phase transition, and therefore, our results are consistent with Refs.~\cite{Pang:2020ilf, Chatziioannou:2019yko, Annala:2019puf, Essick:2023fso, Mroczek:2023zxo}.   
We do not explicitly use information from perturbative QCD~\cite{Komoltsev:2021jzg, Komoltsev:2023zor}, unlike Ref.~\cite{Annala:2019puf}, which argues against EoSs that are hadronic up to very high densities.  
Our analysis is therefore broadly consistent with the existing literature, where such comparisons are possible. 

Finally, we reiterate that our hybrid agnostic-informed formalism can be applied to any nuclear model that can make predictions about the EoS at supranuclear densities.
However, the identification of a breakdown scale is only meaningful if the theory is somewhat trustworthy below some pressure/density scale.  Therefore, constraints from experiments and \emph{ab initio} calculations will be crucial to constrain and validate such theories near saturation density, while astrophysics will be essential for searching for their breakdown at the densities reached in neutron star cores.

\acknowledgments

We thank Thibeau Wouters for helpful comments on this manuscript.
L.B.~and A.H.~thank Mark Alford for useful conversations.  I.L.~thanks Philippe Landry for insightful discussions.
I.L. and K.C. acknowledge support from the Department of Energy under award number DE-SC0023101, and by a grant from the Simons Foundation (MP-SCMPS-00001470).
L.B.~and A.H.~are partly supported by the U.S. Department of Energy, Office of Science, Office of Nuclear Physics, under Award No. $\#$DE-FG02-05ER41375.
A.H.~furthermore acknowledges financial support by the UKRI under the Horizon Europe Guarantee project EP/Z000939/1.  
R.E. is supported by the Natural Sciences \& Engineering Research Council of Canada (NSERC) through a Discovery Grant (RGPIN-2023-03346).

The authors are grateful for computational resources provided by the LIGO Laboratory and supported by National Science Foundation Grants PHY-0757058 and PHY-0823459. This material is based upon work supported by NSF’s LIGO Laboratory which is a major facility fully funded by the National Science Foundation.

\appendix

\section{Constructing the RMFT EoS distribution}
\label{sec:rmft-eos-distribution}

In this Appendix, we discuss how we construct the distribution of RMFT EoSs.
More details are available in Ref.~\cite{Alford:2022bpp}. 
We construct a particular RMFT EoS by making a choice for the RMFT nucleon-meson and meson-meson coupling constants; we choose these constants (which we will here refer to as parameters) so that the resulting EoS is consistent with the properties of isospin-symmetric nuclear matter and neutron matter near saturation.  Since there is uncertainty in the properties of neutron matter, we sample these properties according to \emph{ab initio}, \chiEFT{}, predictions.  We then fit the RMFT parameters using a nonlinear, least-squares procedure, which allows us to produce a set of RMFT parameters which are consistent with the empirical saturation properties of symmetric nuclear matter, and \emph{ab initio} constraints of neutron matter near saturation ($0.5-1.5\,\nsat$).
Therefore, we produce a distribution on the EoS where the only explicit uncertainty comes from uncertainty in the properties of neutron-rich matter near saturation.

We first describe the phenomenological model for symmetric matter, and then the neutron matter constraints, but in practice, the fitting is performed simultaneously.
We fit the binding energy of isospin-symmetric nuclear matter from baryon density $n_B = 0.8\,\nsat$ to $n_B = 1.4\,\nsat$. 
We evaluate the binding energy at 12 density points using the standard empirical power series,
\begin{equation}
    \mathcal{E}(n_B)=(E_B+\frac{\kappa}{2!}\delta^2+\cdots)\,,
    \label{eq:snm_power_series}
\end{equation}
where $\delta\equiv(n_{B}-\nsat)/(3 \nsat)$. 
Here $E_B$ is the binding energy at saturation density and $\kappa$ is the incompressibility of nuclear matter.
The parameter values for isospin-symmetric matter that we fit to are 
\begin{align}
  E_B &= -16\,\text{MeV}\,,\\
  \nsat &= 0.16\,\text{fm}^{-3}\,,\\
  \kappa &= 240\,\text{MeV}\,.
  \end{align} \label{eq:sym-matter-expt}
In contrast with isospin-symmetric matter, neutron matter is not self-bound and therefore cannot be probed in a lab, so we do not have analogous empirical constraints. 
Therefore, instead, we fit to \emph{ab initio} \chiEFT\ calculations of neutron matter~\cite{Tews:2018kmu}. 
The \chiEFT\ calculation we use is based on local \chiEFT\ interactions at N$^2$LO that were constructed in Refs.~\cite{Gezerlis:2013ipa,Gezerlis:2014zia,Tews:2015ufa,Lynn:2015jua}. We choose this \chiEFT\ calculation because, compared to \chiEFT\ calculations at higher order like Refs.~\cite{Drischler:2017wtt, Keller:2020qhx}, Ref.~\cite{Tews:2018kmu} employs a more conservative estimate of the error band. 
The \chiEFT\ results are used over the density range $0.5\,\nsat$ to $1.5\,\nsat$, i.e., above the crust-core transition and where \chiEFT\ is reliable~\cite{Tews:2018kmu,Drischler:2020yad}. 

We sample the \chiEFT\ uncertainty band by creating a set of representative \chiEFT\ neutron matter EoSs using the Gandolfi-Carlson-Reddy (GCR) parametrization~\cite{Gandolfi:2011xu}, which expresses the binding energy per nucleon $\mathcal{E}$ of neutron matter in the form
  \begin{equation}
    {\mathcal{E}}_{\chiEFT}(n_{B}) = a (n_{B}/\nsat)^\alpha + b (n_{B}/\nsat)^\beta\,.
    \label{eq:GCR}
  \end{equation}  
We sample 200,000 EoSs generated from a wide range of GCR parameters $(a,b,\alpha,\beta)$, keeping the 1,109 EoSs that remain completely within the \chiEFT\ uncertainty band between $n_B=0.5\,\nsat$ and $n_B=1.5\, \nsat$. 
For each \chiEFT\ EoS, we fit the RMFT coupling constants, evaluating the \chiEFT\ EoS at 16 density points. A simultaneous fit to these points and the 12 density points of the phenomenological model for symmetric nuclear matter determines the RMFT coupling constants.

The mass-radius relation from the resulting RMFT EoSs is shown in Fig.~\ref{fig:rmf-inj-rmf-posterior}; the multimodality in the RMFT EoSs is driven by a corresponding multimodality in the fit RMFT parameters.  This multimodality is in turn driven by the existence of multiple regions of RMFT parameter space that fit the properties of symmetric and neutron-rich matter comparably well.
For these EoSs we find that the choice of initial guess in the fitting procedure will impact the ``optimal" value for the fit parameters.
In particular, the choice of initial guess for the RMFT parameters we use in this work lead to some  RMFT EoSs ($\sim 8\%$) with $M\geq 2\,M_{\odot}$. 
Using a different set of initial guesses, the multimodality vanishes, but we no longer generate RMFT EoSs with $M\geq 2\,M_{\odot}$. Since we are mainly interested in the range of models that the Lagrangian can produce, and since all RMFT models we generate are consistent with nuclear theory and experiment,  we use the broader (\emph{i.e.} multimodal) distribution.

Since the distribution of RMFT EoSs is influenced by the choice of initial guesses in the fitting procedure, the distribution we present here may not represent the full range of viable models consistent with low-energy nuclear physics for the chosen model Lagrangian.  
Nonetheless, the bulk of fit RMFT EoSs have maximum TOV masses below $2.0\,M_{\odot}$ regardless of the initial guess for the RMFT parameters.  
Therefore, we do not expect the bulk features of the RMFT EoS distribution to change based on the choice of initial guess, though the distribution of EoSs which exceed $2.0\,M_{\odot}$ is likely much more sensitive to this choice.  
Strategies to achieve particular distributions of RMFT EoSs, including direct sampling of RMFT parameters are left to future work.


\section{Efficiently Conditioning high-density GPs on Low-Density Theory}
\label{sec:fix-marginal}

The following is implemented within 
Ref.~\cite{essick_2024_13241272} and was first published there as a technical note alongside the source code. The description is technical, providing a complete record of the procedure of Sec.~\ref{sec:RMF-GP hybrid}.  In particular, we describe here the process necessary for modifying the model-agnostic GP so that at particular pressures it follows the distribution given by the RMFT EoSs.  However, this procedure is generic, and therefore can be applied to modify any GP to strictly obey a particular covariance at particular entries.  

We begin by discussing the particular choice of RMFT-informed and model-agnostic covariance kernels which are used in the construction.  As stated in the main text, both the RMFT-informed and the model-agnostic EoS distributions in fact represent mixture models of Gaussian processes.  The covariance matrices of the RMFT-informed distribution are induced directly by the distribution of RMFT EoSs and therefore have effectively zero parameters (being determined solely by how EoSs are assigned to subgroups to be fit by GPs). The agnostic distribution uses a wide array of possible kernel hyperparameters conditioned on a variety of EoSs, see~\cite{Landry:2018prl, Essick:2019ldf} for details. Different choices made in kernel construction may modify the details of the conclusions drawn here. However, we expect broad consistency because (1) the RMFT GP-EoS distribution is seen to effectively emulate the RMFT EoS distribution, and (2) the model agnostic EoS distribution has been found to have very broad support~\cite{Landry:2018prl,Essick:2019ldf, Essick:2020flb, Essick:2023fso} which would not change under small changes to any of the particular details of the construction.

Assume we have an existing GP that defines a measure for two vectors ($f_a$ and $f_b$ with $N_a$ and $N_b$ elements, respectively) and can be written as
\begin{equation}
    \label{eq:original-process}
    p_\mathrm{O}(f_a, f_b) \sim \mathcal{N}\left( [\mu_a, \mu_b], \begin{bmatrix} C_{aa} & C_{ab} \\ C_{ba} & C_{bb} \end{bmatrix}\right)\,,
\end{equation}
with mean vectors $\mu_a$, $\mu_b$ and covariance matrix $C$ decomposed into $C_{aa}$ ($N_a \times N_a$ elements), $C_{ab}$ ($N_a \times N_b$), $C_{ba}$ ($N_b \times N_a$), and $C_{bb}$ ($N_b \times N_b$).
We note that
\begin{gather}
    C_{aa} = C_{aa}^\mathrm{T}\,, \\
    C_{ab} = C_{ba}^\mathrm{T}\,, \\
    C_{bb} = C_{bb}^\mathrm{T}\,.
\end{gather}
We wish to update this process so that the marginal distribution for $f_b$ follows another process, namely
\begin{equation}
    p_\mathrm{E}(f_b) = \mathcal{N}\left( y_b, \Sigma_{bb} \right)\,,
\end{equation}
while maintaining the rest of the covariance structure encoded in $C$.
We do this by constructing a new process
\begin{equation}
    p_\mathrm{N}(f_a, f_b) = p_\mathrm{O}(f_a|f_b) p_\mathrm{E}(f_b)\,,
\end{equation}
where $p_\mathrm{O}(f_a|f_b)$ can be derived from $p_\mathrm{O}(f_a, f_b)$ in the usual way~\cite{rasmussen2005gaussian} as
\begin{multline}
    p_\mathrm{O}(f_a|f_b) =\\ \mathcal{N}\left( \mu_a + C_{ab} C_{bb}^{-1} (f_b - \mu_b),\ C_{aa} - C_{ab} C_{bb}^{-1} C_{ba} \right)\,.
\end{multline}
Expanding the contractions, grouping like terms, and dropping those that do not depend on either $f_a$ or $f_b$, we obtain a Gaussian in both $f_a$ and $f_b$, and we obtain direct relations for the new mean vectors and (inverse) covariance defined by
\begin{equation}
    p_\mathrm{N}(f_a, f_b) = \mathcal{N}\left([m_a, m_b], \begin{bmatrix} \Gamma_{aa} & \Gamma_{ab} \\ \Gamma_{ba} & \Gamma_{bb} \end{bmatrix}^{-1} \right)\,,
\end{equation}
as follows:
\begin{align}
    \Gamma_{aa} & = \left(C_{aa} - C_{ab} C_{bb}^{-1} C_{ba}\right)^{-1}\,, \\
    \Gamma_{ab} & = - \left(C_{aa} - C_{ab} C_{bb}^{-1} C_{ba}\right)^{-1} C_{ab} C_{bb}^{-1}\,, \\
    \Gamma_{bb} & = C_{bb}^{-1} C_{ba} \left(C_{aa} - C_{ab} C_{bb}^{-1} C_{ba}\right)^{-1} C_{ab} C_{bb}^{-1} + \Sigma_{bb}^{-1}\,,
\end{align}
and
\begin{align}
    m_a & = \mu_a + C_{ab} C_{bb}^{-1} \left( y_b - \mu_b \right)\,, \\
    m_b & = y_b\,.
\end{align}
Finally, we can solve for
\begin{equation}
    \gamma = \Gamma^{-1} = \begin{bmatrix} \gamma_{aa} & \gamma_{ab} \\ \gamma_{ba} & \gamma_{bb} \end{bmatrix}\,,
\end{equation}
by recognizing that
\begin{equation}
    \begin{bmatrix} \gamma_{aa} & \gamma_{ab} \\ \gamma_{ba} & \gamma_{bb} \end{bmatrix} \begin{bmatrix} \Gamma_{aa} & \Gamma_{ab} \\ \Gamma_{ba} & \Gamma_{bb} \end{bmatrix} = \mathbb{I}\,.
\end{equation}
Further simplification yields
\begin{align}
    \gamma_{aa} & = \left[\Gamma_{aa} - \Gamma_{ab} \Gamma_{bb}^{-1} \Gamma_{ba} \right]^{-1}\,, \\
    \gamma_{ab} & = C_{ab} C_{bb}^{-1} \Sigma_{bb}\,, \\
    \gamma_{ba} & = \Sigma_{bb} C_{bb}^{-1} C_{ba}\,, \\
    \gamma_{bb} & = \Sigma_{bb}\,,
\end{align}
where we have left $\gamma_{aa}$ in terms of $\Gamma$ because of the length of the expression but have substituted and simplified the rest of the terms.
Note that the marginal distribution $p_\mathrm{N}(f_b) = \mathcal{N}(y_b,\Sigma_{bb}) = p_\mathrm{E}(f_b)$, as desired. On the other hand, the distribution of $y_a$ is modified from it's original form in  Eq.~\eqref{eq:original-process}.  In our case, this is an indication that the conditioning process modifies the EoS distribution even at pressures which we did not explicitly require the distribution to follow RMFT.


\subsection{Modifications for numerical stability}
\label{sec:numerical stability}

In general, $m_a$ and $\gamma_{aa}$ can suffer from issues associated with numerical stability.
This is because they involve the inversion of (possibly) high-dimensional matrices that may be ill-conditioned.
While the preceding is exact, we therefore implement two additional approximations to help better control the calculations.


\subsubsection{Damping $C_{ab}$, $C_{ba}$, and $C_{bb}$ to make them easier to invert}

One issue we have found is that strong correlations in $C_{bb}$ can make numerical inversion difficult.
Given that we wish to replace $C_{bb}$ with $\Sigma_{bb}$ anyway, and really only wish there to be a relatively smooth transition between $f_b$ and $f_a$, we modify $C_{ab}$, $C_{ba}$, and $C_{bb}$ in order to damp the off-diagonal elements (and therefore make them easier to invert).

Specifically, we define a squared-exponential damping term
\begin{equation}
    D(x_i, x_j) = \exp\left(-\frac{(x_i - x_j)^2}{l^2}\right)\,,
\end{equation}
and a white noise contribution that modify $C$ so that
\begin{gather}
    (C_{ab})_{ij} \rightarrow (C_{ab})_{ij} D(x_i, x_j)\,, \\
    (C_{bb})_{ij} \rightarrow (C_{bb})_{ij} D(x_i, x_j) + \sigma_\mathrm{W}^2 \delta_{ij}\,.
\end{gather}
We then use these modified $C_{ab}$ and $C_{bb}$ within the expressions in the previous section.

This modifies the original process, but as long as $l$ is relatively large and $\sigma_\mathrm{W}$ is relatively small, the modifications will be minor over the transition between $f_b$ and $f_a$.
Empirically, we find that $l = 5.0$ and $\sigma_\mathrm{W} = 0.01$ work well when updating our our model-agnostic priors.


\subsubsection{Approximation for $\gamma_{aa}$ when $\Sigma_{bb}$ is small}

Finally, it will often be the case that $\Sigma_{bb}$ will be much smaller than $C_{bb}$ (with respect to an appropriate matrix norm).
That is, we wish to update a process to restrict the marginals of certain covariates to be more tightly constrained than they otherwise would be.

By repeated use of the approximation
\begin{equation}
    (A + X)^{-1} \approx A^{-1} - A^{-1} X A^{-1}\,,
\end{equation}
we can see that this limit corresponds to
\begin{equation}
    \Gamma_{bb}^{-1} \approx \Sigma_{bb} - \Sigma_{bb} C_{bb}^{-1} C_{ba} \left( C_{aa} - C_{ab} C_{bb}^{-1} C_{ba} \right)^{-1} C_{ba} C_{bb}^{-1} \Sigma_{bb}\,,
\end{equation}
and (retaining terms linear in $\Sigma_{bb}$)
\begin{align}
    \gamma_{aa}
        & \approx C_{aa} - C_{ab} C_{bb}^{-1} C_{ba} + C_{ab} C_{bb}^{-1} \Gamma_{bb}^{-1} C_{bb}^{-1} C_{ba} \nonumber \\
        & \approx C_{aa} - C_{ab} C_{bb}^{-1} C_{ba} + C_{ab} C_{bb}^{-1} \Sigma_{bb} C_{bb}^{-1} C_{ba}\nonumber\\
        & \approx C_{aa} - C_{ab} C_{bb}^{-1} (C_{bb} - \Sigma_{bb}) C_{bb}^{-1} C_{ba}\,.
\end{align}
This makes sense in two limiting cases
\begin{itemize}
    \item $\Sigma_{bb}=0$ : we know $f_b$ exactly and obtain the standard expression for the covariance for $f_a|f_b$,
    \item $\Sigma_{bb}=C_{bb}$: we do not update the original process, and as such we obtain $\gamma_{aa} = C_{aa}$.
\end{itemize}

Finally, we offer one more interpretation of this expression.
If we considered the standard expression for $f_a|f_b$ with some covariance for $f_b$, say $\mathcal{C}_{bb}$, we would obtain
\begin{equation}
    \gamma_{aa} = C_{aa} - C_{ab} \mathcal{C}_{bb}^{-1} C_{ba}\,,
\end{equation}
and therefore, by matching this to our approximation, we see that
\begin{align}
    \mathcal{C}_{bb}
        & = C_{bb} \left( C_{bb} - \Sigma_{bb} \right)^{-1} C_{bb} \nonumber \\
        & \approx C_{bb} \left( C_{bb}^{-1} + C_{bb}^{-1} \Sigma_{bb} C_{bb}^{-1} \right) C_{bb} = C_{bb} + \Sigma_{bb}\,.
\end{align}
In this limit, then, we can interpret updating the marginal distribution as equivalent to the standard procedure of conditioning the process for $f_a$ on a noisy observation of $f_b$ with observed values $y_b$ and measurement uncertainty $\Sigma_{bb}$.
Historically, this is what was actually done in Refs.~\cite{Essick:2020flb, Essick:2021kjb, Essick:2021ezp}, and we now see why it provided a decent approximation.


\section{Inferring a breakdown with gravitational-wave signals and pulsar timing measurements alone}
\label{sec:simulated-inference-gws}
In this Appendix, we consider the question of how many gravitational-wave observations (when paired with existing heavy pulsar mass measurements) will be required to determine the breakdown of an RMFT.  
We do this to explicitly evaluate the prospect of developments of future gravitational-wave detectors.  With next generation detectors such as Cosmic Explorer~\cite{Evans:2023euw, Gupta:2023lga}, or Einstein Telescope~\cite{Abac:2025saz}, in principle hundreds to thousands of informative binary neutron star mergers could be identified within a handful of years of operation~\cite{Evans:2023euw, Abac:2025saz}.
In contrast, there are a limited number of plausible x-ray timing targets with which to constrain the EoS~\cite{Bogdanov:2019ixe}.

We start with more details on the gravitational-wave injections. 
We use a binary population that is consistent with observations of merging neutron stars, sampling 50 sources from a representative population model which is uniform in component masses from 1.0 to $2.0\, M_{\odot}$.\footnote{This is less than the TOV maximum mass for the EoSs we use for simulated data, assuming that the population of merging NSs is not limited by the TOV maximum mass; see e.g., Ref.~\cite{Golomb:2024lds} for a discussion.  We assume this because high-mass neutron star mergers are essentially uninformative with respect to the EoS, and more importantly, high mass NSs are generally not distinguishable from black holes anyway in gravitational waves~\cite{Farah:2021qom, Fishbach:2020ryj, Essick:2020ghc}}  We simulate signals in the Hanford and Livingston detectors~\cite{LIGOScientific:2014pky} corresponding to the sampled sources into Gaussian noise at the level of A+ detector sensitivity \cite{a_plus_sensitivity}, placing the sources uniformly in comoving volume from 1 to 300\,Mpc, which we expect to produce a substantial fraction of ``informative" GW signals. 
We assume the neutron stars are spinning slowly, with dimensionless component spins isotropically distributed with magnitudes $|\vec \chi| < 0.05$.  
We analyze all sources that are recovered with optimal signal-to-noise ratio (SNR) $> 10.0$ using the \texttt{bilby} parameter estimation code~\cite{Ashton2019} using the \texttt{IMRPhenomPV2\_NRTidalv2} waveform~\cite{Dietrich:2019kaq}.\footnote{In general, selecting events based on the optimal SNR can induce biases~\cite{Essick:2023upv}. However, we do not change the mass, distance, or spin distribution, and the EoS contributes negligibly to selection effects.}
We then downsample the total set of analyzed sources in order to produce variable catalog sizes.

\begin{figure}
    \centering
    \includegraphics[width=0.49\textwidth]{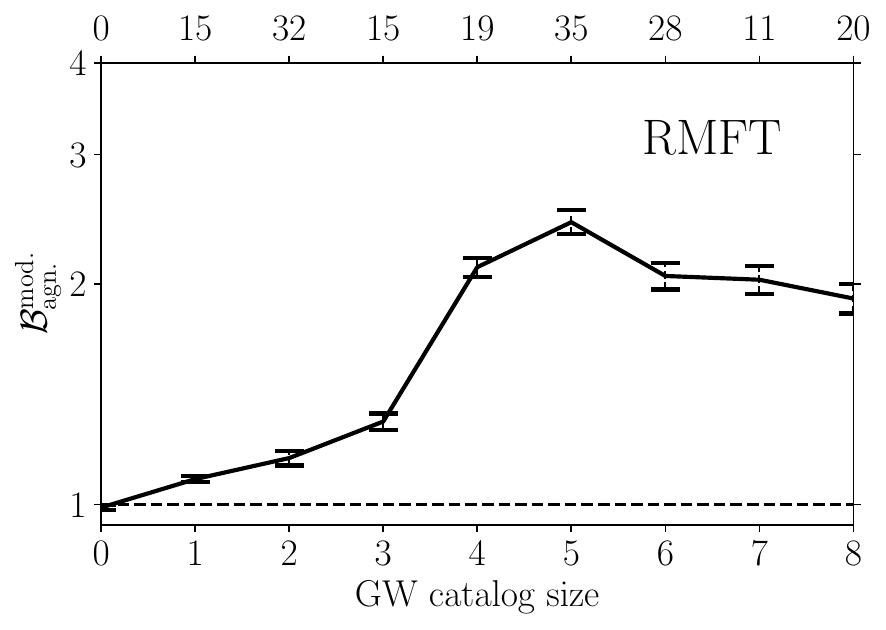}
    \includegraphics[width=0.49\textwidth]{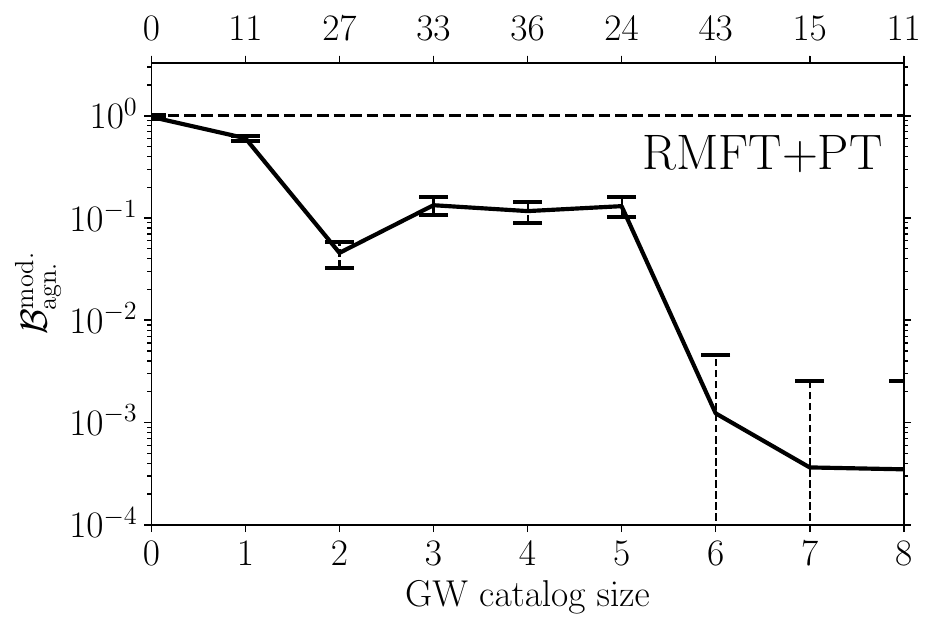}
    \caption{The ratio of evidences (i.e. Bayes factor), between models assuming a transition from an RMFT informed model at $p/c^2 = 10^{14}\,\rm{g/cm^3}$ and at  $p/c^2 =  3\times 10^{13}\,\rm{g/cm^3}$. The top panel is for the case of an RMFT injection, whereas the bottom panel is for the case of an RMFT that transitions to a constant speed-of-sound EoS. We plot the Bayes Factor as a function of the number of events analyzed. On the top x-axis of each figure, we include the SNR of the event added to the catalog.}
    \label{fig:evidence-gws-only}
\end{figure}

In Fig.~\ref{fig:evidence-gws-only}, we show the Bayes factor between models transitioning from RMFT-informed to model-agnostic priors at $p/c^2 = 10^{14}\, \rm{g}/\rm{cm}^3$ and $3\times 10^{13}\, \rm{g}/\rm{cm}^3$, respectively.
We consider two cases: inferring an RMFT EoS (top panel, see Sec.~\ref{sec:rmf-recovery}) and an RMFT EoS which has a phase transition to a constant speed of sound (bottom panel, see Sec.~\ref{sec:pt-recovery}) at $p/c^2 = 2.3 \times 10^{13}\, \rm{g}/\rm{cm}^3$.  
As in the main text, we find no strong evidence either for or against in the case of an RMFT injection. 
In this case, all models can describe the data, so the Bayes factors essentially reflect the priors.  
In the case of the phase transition, we find that meaningful evidence ($\mathcal B \lesssim 10^{-3}
$) against an RMFT description up to high pressure after $\sim 7$ events,\footnote{This is a plausible estimate for the number of detections of neutron star binaries with LIGO A+ sensitivity given current rate estimates~\cite{KAGRA:2021duu}.  Future detectors will produce far more detections, which will lead to sampling challenges~\cite{Essick:2023fso}, and far louder detections~\cite{Chatziioannou:2021tdi}, which will lead to systematic biases due to waveform mismodeling~\cite{Read:2023hkv}.  Collectively these considerations make projection with next-generation detectors using our methodology technically challenging, we therefore restrict to A+.} with most of the evidence accruing from a handful of informative events.    
For reference, the mass of the lowest mass neutron star with a core undergoing the phase transition (the ``transition mass") is $\sim 0.6\, M_{\odot}$, well below any of the simulated neutron stars.  
Therefore all simulated neutron stars have substantial quark cores. 
The highest SNR event in the phase transition EoS case is also the most informative (SNR of 43, 6th event in the catalog).
However, certain events even with large SNR ($\gtrsim 30$) are not necessarily informative, usually because their masses are too large and therefore lack a measurable tidal signature.
Nonetheless, these results indicate that upcoming gravitational wave data could lead to stronger constraints on the RMFT breakdown scale.

\section{Additional analysis with a weaker phase transition}
\label{sec:additional-analysis}

\begin{figure}
    \centering
    \includegraphics[width=0.99\linewidth]{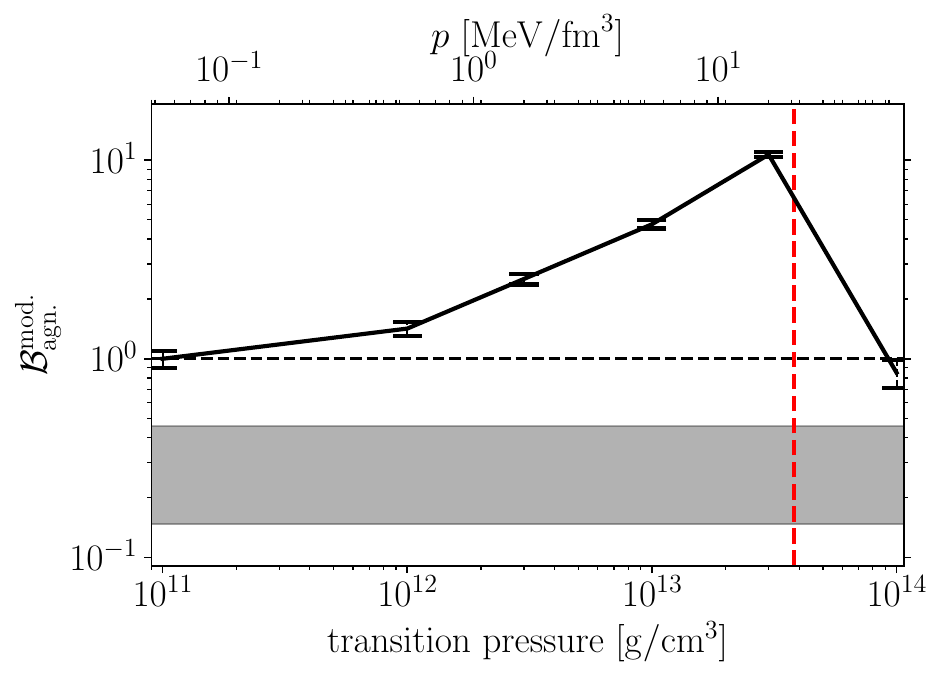}
    \caption{Same as Fig.~\ref{fig:rmf-pt-bayes-factors}, but for a different choice of phase transition parameters.}
    \label{fig:other-pt-evidence}
\end{figure}

\begin{figure}
    \centering
    \includegraphics[width=0.99\linewidth]{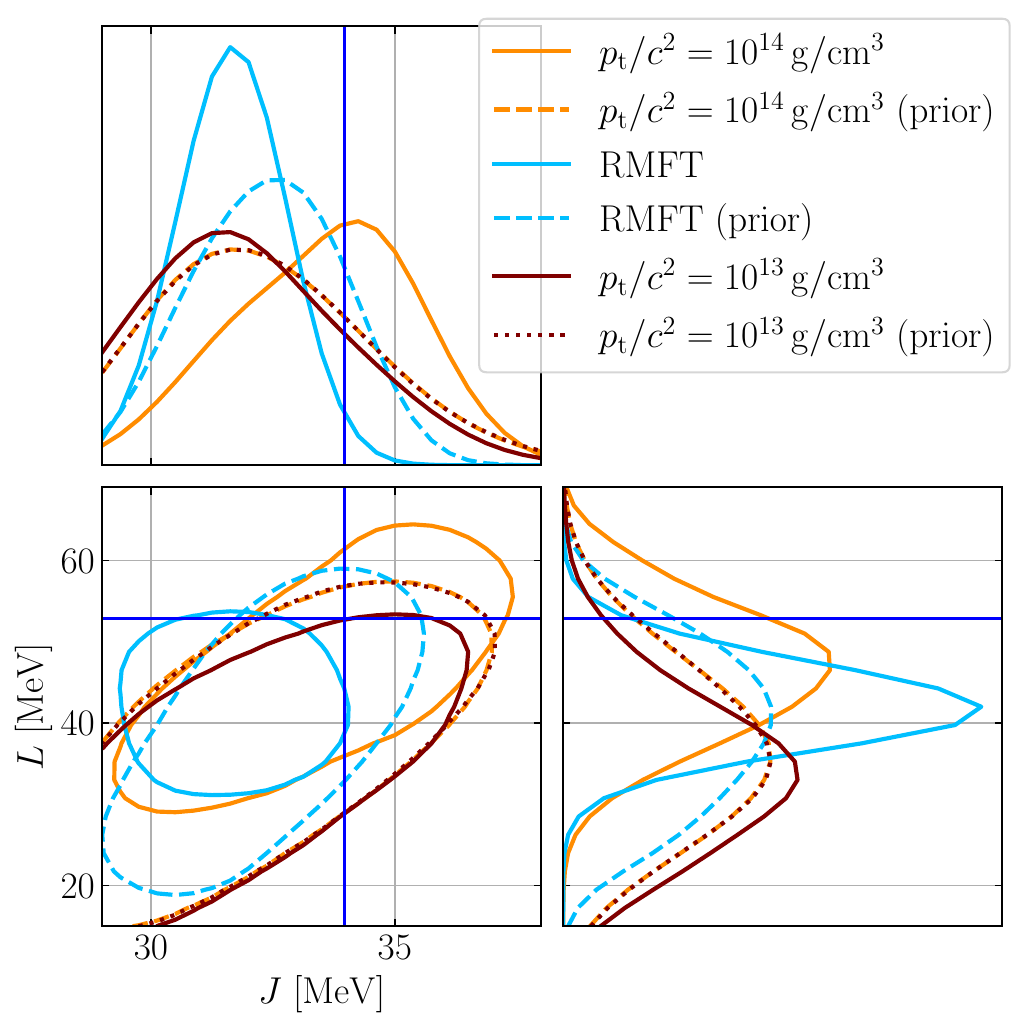}
    \caption{Same as Fig.~\ref{fig:rmf-pt-symmetry-params}, but for a different choice of phase transition parameters.}
    \label{fig:other-pt-sym-params}
\end{figure}

We repeat our simulation analysis with the same RMFT EoS at low densities, but with a transition to a $c_s^2 = 0.7$ EoS at $2\nsat$, with a $\Delta \varepsilon/ \varepsilon$ of 0.3.  The resulting EoS much more closely masquerades as an RMFT EoS.  We show the analogous plot to Fig.~\ref{fig:rmf-pt-bayes-factors}
 in Fig.~\ref{fig:other-pt-evidence}.  We find that such a change to the simulation EoS phase transition parameters substantially reduces the evidence against the RMFT model, however,  we find the general trend seen in Fig.~\ref{fig:rmf-pt-bayes-factors} still holds.   Further, we show in Fig.~\ref{fig:other-pt-sym-params} that when using the RMFT EoS set, the incorrect symmetry parameters are identified at the 90\% level for this simulation EoS.  Therefore, while astrophysical observations alone may struggle to distinguish such an observation, the additional inclusion of nuclear data from experiments and \emph{ab initio} calculations should nonetheless allow the confident exclusion of the RMFT holding up to arbitrarily high densities.

\clearpage
\bibliography{references}

\end{document}